\documentclass[a4paper,12pt]{iopart}
\usepackage{iopams}
\usepackage{bbm}
\usepackage{graphicx}
\usepackage{amssymb}
\usepackage{hyperref}

\begin{document}

%\title[]{Quantum state transfer via thermal phonon waveguides}

%\title[]{Quantum networks based on thermal phonon waveguides}

\title[]{Continuous mode cooling and phonon routers for phononic quantum networks}

\author{S.~J.~M.~Habraken$^1$, K.~Stannigel$^{1,2}$,  M.~D.~Lukin$^3$, P.~Zoller$^{1,2}$ and P.~Rabl$^1$}

\address{$^1$Institute for Quantum Optics and Quantum Information of the Austrian Academy of Sciences, Innsbruck, Austria}
\address{$^2$Institute for Theoretical Physics, University of Innsbruck, Austria}
\address{$^3$Department of Physics, Harvard University, Cambridge MA, USA}
\ead{steven.habraken@uibk.ac.at}

\date{\today}

\begin{abstract}
We study the implementation of quantum state transfer protocols in phonon networks, where in analogy to optical networks, quantum information is transmitted through propagating phonons in extended mechanical resonator arrays or phonon waveguides. We describe how the problem of a non-vanishing thermal occupation of the phononic quantum channel can be overcome by implementing optomechanical multi- and continuous mode cooling schemes to create a `cold' frequency window for transmitting quantum states. In addition, we discuss the implementation of phonon circulators and switchable phonon routers,  which rely on strong coherent  optomechanical interactions only, and do not require strong magnetic fields or specific materials. Both techniques can be applied and adapted to various physical implementations, where phonons coupled to spin or charge based qubits are used for on-chip networking applications.
\end{abstract}

\maketitle

\section{Introduction}

The successful application of laser cooling techniques for cooling isolated vibrational modes of micro- and nanomechanical devices \cite{Metzger2004,Gigan2006,Arcizet2006,Kleckner2006,Corbitt2007,Thompson2008,Schliesser2008,Wilson2009} has recently attracted a lot of interest in the control of macroscopic phononic degrees of freedom on a single quantum level.
By now the preparation of mechanical resonators close to the quantum ground state 
has been achieved in different experimental settings \cite{OConnell2010,Teufel2011,Chan2011} and coherent interfaces between mechanics and other quantum systems like superconducting qubits \cite{OConnell2010,LaHaye2009}, 
spins \cite{Arcizet2011,Kolkowitz2012} 
or photons \cite{Fiore2011,Verhagen2012} are currently developed. Beyond new possibilities to address fundamental questions in quantum physics \cite{Armour2002,Marshall2003,Romero-Isart2012,Pikovski2012},  these experimental developments also provide the foundation for new, phonon-based quantum technologies.  
For example, in the context of quantum information processing and quantum communication, optomechanical (OM) slowing of light \cite{Weis2010,Safavi-NaeiniNature2011} and first steps towards realizing a mechanical quantum memory \cite{Fiore2011,Verhagen2012}
have been demonstrated, and the use of mechanical quantum transducers for interfacing different qubit systems \cite{Rabl2010,Stannigel2010,Safavi-NaeiniNJP2011} has been proposed. For these applications mechanical systems benefit from the ability to interact with a wide range of other electric, magnetic and optical quantum systems, while still maintaining long coherence times and being compatible with scalable nano-fabrication techniques.

The use of phonons for quantum information science applications is not new. Already in the first proposals for quantum computers, it has been suggested to employ  vibrational modes of a trapped ion Wigner crystal for transmitting quantum information between spatially separated qubits \cite{Cirac1995}. More recently it has been shown that these ideas could equally well be applied in systems of coupled macroscopic mechanical resonators \cite{Rabl2010,Eisert2004,Schmidt2012}, which extends the concept of a mechanical quantum bus to a wider range of atomic and solid-state systems. 
In analogy to optical fields,  phonons can be confined in phonon cavities (e.g. represented by a high $Q$ mechanical resonator), but also propagate freely along phononic waveguides. This suggests that many quantum communication and state-transfer protocols developed in the context of optical quantum networks \cite{Cirac1997,Kimble2008,Ritter2012} could -- on a smaller physical scale -- also be implemented using acoustic phonons. Here, new approaches to design and pattern phonon waveguides based on phononic crystals structures \cite{Olsson2009,Safavi-Naeini2010} provide a very promising and versatile platform for realizing such phonon networks in practice.  However, compared with the relatively advanced field of optical quantum networks \cite{Kimble2008}, many equivalent control techniques still have to be developed for phononic quantum systems, which face the additional challenge that  thermal noise in phononic channels is not negligible and would usually by far exceed quantum signals encoded in a single phononic excitation.

In this work we address the problem of implementing quantum communication protocols in thermal phononic channels and show, how the addition of OM control elements can be used to realize a faithful transfer of quantum information between different nodes of the network. As a first key element to achieve this task, we describe the generalization of OM laser cooling techniques to multi- and continuous mode setups. This approach is motivated by a recent proposal for interfacing OM systems (OMS) with phonon waveguides \cite{Safavi-NaeiniNJP2011}, and leads to a strong suppression of thermal noise within the relevant transmission bandwidth. Therefore, instead of pursuing the otherwise challenging task of cooling the whole network,  this technique creates a `cold' frequency window, which is sufficient to coherently  transfer single quanta through an otherwise `hot' phononic channel. 

As a second control tool we describe the implementation of phonon circulators and switchable phonon routers, which enable a directed transfer of propagating phonons through large 1D or 2D networks. In the optical domain circulators or other non-reciprocal devices are usually based on the Faraday-effect. In principle similar effects also exist for acoustic phonons in certain materials \cite{Luethi2005}. However, due to the required large magnetic fields and the use of materials with non-optimized mechanical properties, this approach is not suited  for on-chip phonon quantum networks.  Instead, we propose an integrated circulator for acoustic phonons that relies solely on an OM induced non-reciprocity~\cite{Manipatruni2009,Hafezi2012}, where the directionality is imposed by the phase relation between two optical driving fields.  Thereby, the device can be switched on or reversed conveniently and in combination with the above-mentioned cooling techniques, this coherent OM routing scheme is in principle sufficient to fully control  the distribution of quantum information in large-scale phonon networks.

The remainder of this paper is structured as follows.  In \sref{sec:PhononNetworks} we first present a brief introduction to phonon networks and the general input-output formalism, which is used to model these systems. In \sref{sec:NoiseEater} we revisit OM laser cooling and generalize it to multi- and continuous mode scenarios. As a basic application we then discuss in \sref{sec:StateTransfer} how an OM noise filter can be applied to transmit a quantum state through a thermal   
channel. In \sref{sec:PhononRouter} we describe the realization of phonon circulators and routers using coherent OM interactions. Finally, in \sref{sec:Implementations} we outline several potential systems for implementing phonon networks and then summarize the main results and conclusions of this work in \sref{sec:Conclusions}. 

\section{Phonon quantum networks}\label{sec:PhononNetworks}

\begin{figure}[t]
\begin{center}
\includegraphics[width=0.85\linewidth]{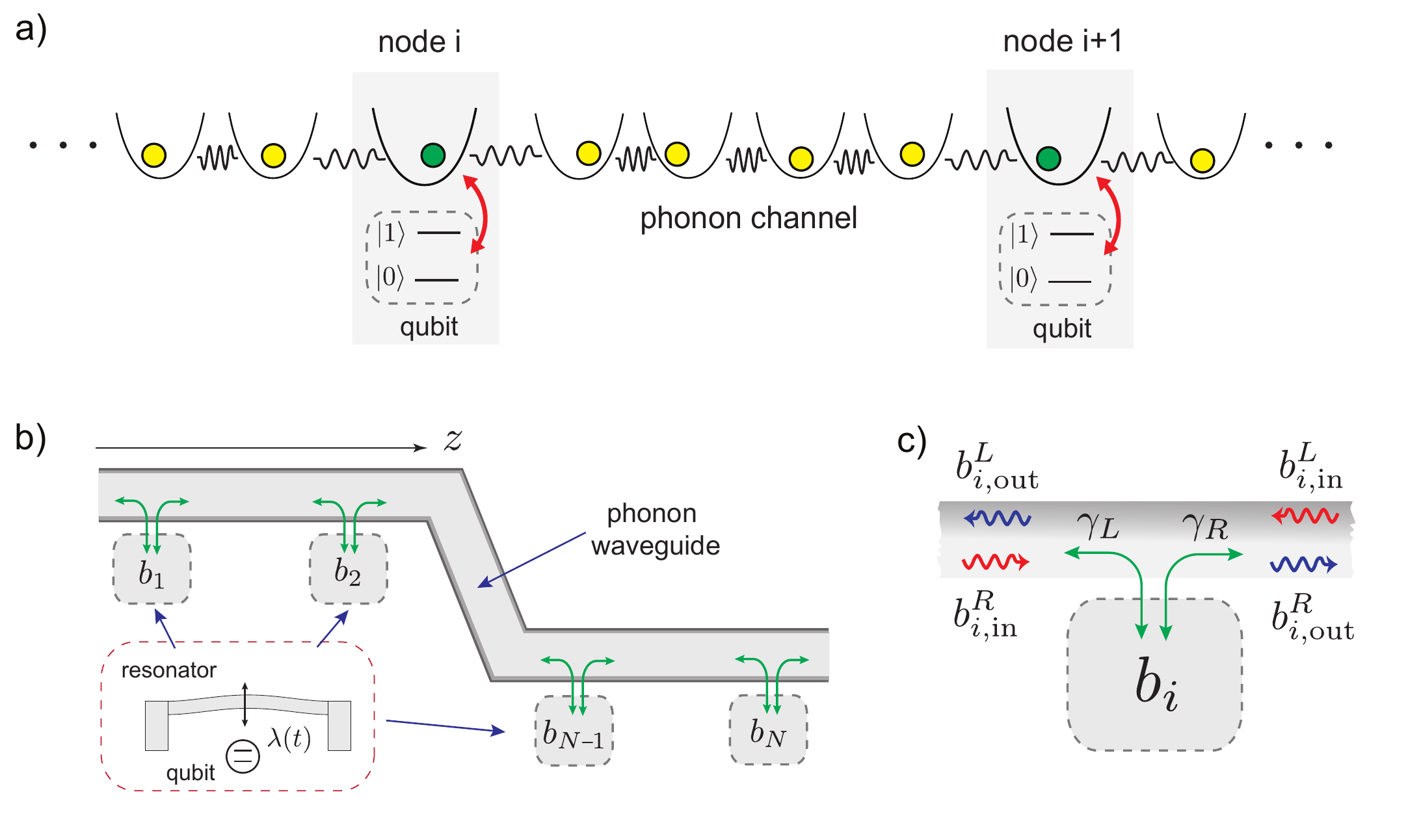}
\caption{a) Schematic representation of a generic phonon quantum network. At each node qubits interact with a localized phonon mode, which in turn is weakly coupled to a phononic channel represented by an extended array of coupled mechanical resonators or a 1D phonon waveguide. b) A phonon quantum network consisting  of a set of localized vibrational modes, which are side-coupled to a common phonon waveguide and mediate the interaction between qubits and the propagating phonon modes (For a specific implementation of such a setup see reference~\cite{Safavi-NaeiniNJP2011}). c) The localized phonons decay into left- and right propagating modes in the waveguide with corresponding rates $\gamma_L$ and $\gamma_R$.  The $b^{L/R}_{i,{\rm in}}$ and $ b^{L/R}_{i,{\rm out}}$ denote the incident and outgoing waveguide fields at each node. See main text for more details.}
\label{fig:PhononNetwork}
\end{center}
\end{figure}

\Fref{fig:PhononNetwork}a shows a schematic representation of a generic phonon quantum network, where individual nodes $i=1,\dots, N$ are connected via a mechanical quantum bus. In analogy to optical quantum networks, each node contains  an isolated solid state two-level system (`qubit') with quantum information encoded in states $|0\rangle$ and $|1\rangle$. The qubits interact with localized phonon modes (`phonon cavities'), which are in turn weakly coupled to a common set of propagating modes  of a phonon waveguide or coupled mechanical resonator array (`quantum channel'). The full Hamiltonian of this system is given by  
\begin{equation}
H= \sum_{i=1}^N H_{\rm node}^i +H_{\rm channel} +H_{\rm int}, 
\end{equation}
where $H_{\rm node}^i$ and $H_{\rm channel}$ describe the  dynamics of the individual nodes and the phononic channel, respectively, and $H_{\rm int}$ accounts for the coupling between them. In the following, we assume that the Hamiltonian for the individual nodes takes the form $(\hbar=1)$
\begin{equation}\label{eq:Hnode}
H_{\rm node}^i(t) =  \omega_m b_i^\dag b_i  + \frac{\Delta^i_q(t)}{2}\sigma^i_{z}+\lambda_i(t)\left(\sigma^i_{+}b_i+\sigma^i_{-}b_i^{\dag}\right)\;,
\end{equation}
where the $\sigma_k^i$ are Pauli operators and $b_i$ the bosonic operators for the local mechanical modes of frequency $\omega_m$. The third term describes the coupling between the qubits and the local mechanical modes. For each node the qubit  frequency splitting $\Delta^i_q(t)$ and the qubit-resonator coupling $\lambda_i(t)$ can be tuned independently by changing the qubit frequency or modulating the coupling with local control fields (see section \sref{sec:Implementations}).  For $\omega_m\sim \Delta^i_q$ this allows for a controlled mapping of the qubit state onto a phonon superposition, which we will use in our discussion of the state transfer below.  The implementation of qubit-resonator interactions as given in equation \eref{eq:Hnode} has been described for various charge- and spin-based qubits in the literature and in \sref{sec:Implementations} we briefly summarize some of the most relevant systems. 

\subsection{Phonon channels}
The mechanical  quantum channel which is used to communicate between the nodes can in general be represented by a chain of $N_{\rm ch}$ coupled mechanical resonators, 
\begin{equation}
H_{\rm channel}= \sum_{\ell=1}^{N_{\rm ch}}  \frac{p_\ell^2}{2m} +\frac{1}{2} m \omega_0^2 x_\ell^2 + \frac{k}{2}\sum_{\ell=1}^{N_{\rm ch}-1}  (x_\ell-x_{\ell+1})^2\;.
\end{equation}
Here $x_\ell$ and $p_\ell$ are the position and momentum operators, $m$ is the effective mass, $\omega_0$ the bare oscillation frequency  and the spring constant $k$ accounts for the  nearest-neighbor coupling. For small $N_{\rm ch}$  the coupled resonators form a discrete set of collective eigenmodes. In this case quantum state transfer and quantum information processing protocols between two qubits can be implemented by addressing only a single collective mode, as has been discussed in the context of trapped ion systems \cite{Cirac1995} or coupled nanomechanical \cite{Rabl2010} and optomechanical \cite{Schmidt2012} resonators.  

In this work we are primarily interested in the opposite regime of extended arrays, $N_{\rm ch}\gg 1$. In this limit  $H_{\rm channel}$ can be represented by a dense set of plane wave modes, $H_{\rm channel} = \sum_q  \omega_q b_q^\dag b_q$, where $[b_q,b_{q'}^\dag]=\delta_{q,q'}$,  $\omega_q$ is the phonon dispersion relation, and, for a lattice spacing $a$, the momentum label $q$ is restricted to the first Brillouin zone $q\in (-\frac{\pi}{a},\frac{\pi}{a}]$. This scenario is realized, for example, in large arrays of coupled nanomechanical beams \cite{Buks2002} or in phononic band gap structures \cite{Olsson2009}, where each resonator $\ell$ corresponds to a vibrational mode of a unit cell, typically of size $a\sim 1\,\mu$m. In the continuum limit we can identify a frequency range $\Delta \omega$ away from the band edges in which the coupled resonator array exhibits an approximately linear dispersion $\omega_q\approx  \tilde \omega_0 + c |q|$, where $c$ is the effective speed of sound and $\tilde \omega_0$ a frequency offset. 
In this case it is convenient to introduce the normalized displacement field $\Phi(z)= \Phi_R(z) + \Phi_L(z)$, where 
\begin{equation}\label{eq:PhiLR}
\Phi_{R} (z) =\sqrt{\frac{2c}{L}} \sum_{q>0}  e^{ iq z} b_q, \qquad   \Phi_{L} (z)=\sqrt{\frac{2c}{L}} \sum_{q>0}  e^{- iq z} b_{-q},
\end{equation}
describe right- and left-moving mechanical excitations of the phononic channel. Under the assumption of a linear dispersion  these fields obey $[ \Phi_{R}(z,t), \Phi^\dag_L(z',t')]= 0$ and   
\begin{equation}\label{eq:PhiLRCommutator}
\left[ \Phi_{R/L}(z,t), \Phi^\dag_{R/L}(z',t')\right] \simeq  e^{-i\tilde \omega_0 (t-t')}  \delta\left( t-t' \mp (z-z')/c\right).
\end{equation}  
By assuming that the frequency of the local mechanical modes $\omega_{m}$ also lies within this frequency range, the coupling between the local resonators and the waveguide modes can be approximated by 
\begin{equation}\label{eq:Hint}
H_{\rm int}\simeq \sqrt{\frac{\gamma}{2}} \int_0^L dz\, \sum_{i=1}^N  \delta(z-z_i)   \left( b_i^\dag \Phi(z) + b_i \Phi^\dag(z)\right)\;,
\end{equation}
where $L$ is the length of the waveguide and $z_i$ are the positions of the nodes along the waveguide (see \fref{fig:PhononNetwork}b). Below we identify $\gamma$ as the total decay rate of the local resonator modes into the waveguide.  Note that equations \eref{eq:PhiLRCommutator} and \eref{eq:Hint} are valid for times $|t-t'|>\Delta \omega^{-1}$ and distances $|z-z'|> a$. An explicit and more detailed derivation of these results is given in \ref{app:ResonatorArray}  for the case of a simple coupled resonator chain. 

\subsection{Input- output relations}
 
Under the validity of equations \eref{eq:PhiLRCommutator} and \eref{eq:Hint}  and in the limit where $\omega_m$ and the bandwidth $\Delta \omega $ are large compared to the other characteristic frequency scales, we can eliminate the waveguide modes and use an input- output formalism \cite{QuantumNoise} to describe the effective dynamics of the coupled nodes. Using \eref{eq:PhiLRCommutator} and \eref{eq:Hint} and making a standard Born-Markov approximation,  we can derive  a set of coupled quantum Langevin equations (QLEs) \cite{QuantumNoise}. For each node we obtain
\begin{equation}\label{eq:QLEbi}
\dot b_i= i [H_{\rm node},b_i] - \frac{\gamma}{2}  b_i  - \sqrt{\gamma_R }b^R_{i,{\rm in}}(t)- \sqrt{\gamma_L }b^L_{i,{\rm in}}(t)\;,
\end{equation}
where $\gamma=\gamma_R+\gamma_L$ is the total decay rate of phonons into the waveguide.  For side-coupled phonon cavities we would usually have $\gamma_R=\gamma_L$, but we below we describe scenarios where the emission either to the left or to the right is effectively switched off.   
In equation \eref{eq:QLEbi} we have defined  incoming fields (see \fref{fig:PhononNetwork}c)
\begin{equation} 
b^L_{i,{\rm in}}= \lim_{\epsilon \rightarrow 0} \Phi_L(z_i+\epsilon),\qquad  b^R_{i,{\rm in}}= \lim_{\epsilon \rightarrow 0} \Phi_R(z_i-\epsilon),
\end{equation}
which specify the left and right moving waveguide modes before they interact with $b_i$. Similarly, we introduce the scattered outgoing fields
 \begin{equation} 
b^L_{i,{\rm out}}= \lim_{\epsilon \rightarrow 0} \Phi_L(z_i-\epsilon),\qquad  b^R_{i,{\rm out}}= \lim_{\epsilon \rightarrow 0} \Phi_R(z_i+\epsilon)\;.
\end{equation}
The dynamics of the whole network is then described by the QLEs \eref{eq:QLEbi} together with the input-output-relations at each node,
\begin{equation}
b^{R,L}_{i, {\rm out}}(t)=  b^{R,L}_{i,{\rm in}}(t) +\sqrt{\gamma_{R,L}} b_i(t),
\end{equation}
and the free propagation 
\begin{equation}
b^R_{i, {\rm in}}(t)=  b^R_{i-1,{\rm out}}(t-\tau_{i,i+1}),\qquad  b^L_{i, {\rm in}}(t)=  b^L_{i+1,{\rm out}}(t-\tau_{i,i+1}),
\end{equation}
where $\tau_{i,i+1} = |z_{i+1}-z_i|/c$. Under realistic conditions we must also account for intrinsic phonon losses, backscattering and rethermalization, which we address below. 

\subsection{Optomechanical control techniques for phonon networks}

\begin{figure}[t]
\begin{center}
\includegraphics[width=0.8\linewidth]{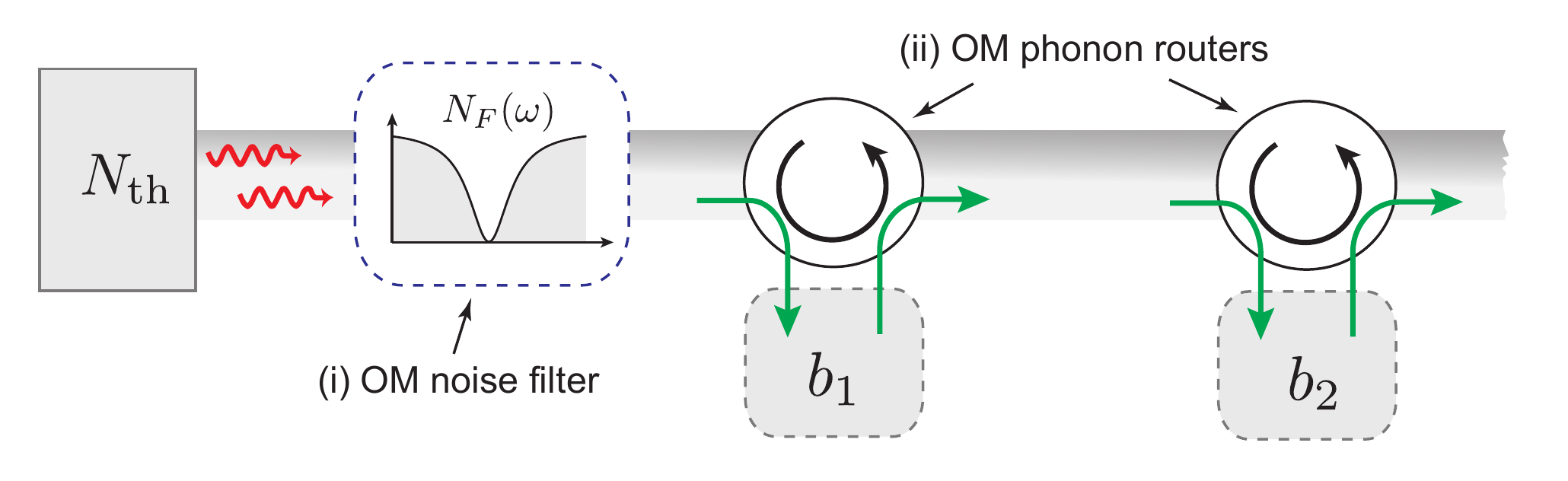}
\caption{Basic setup illustrating the application of OM control techniques for implementing a directed quantum state transfer between two nodes through a thermal phonon network. (i) An OM noise filter is used to suppress incident thermal noise $\sim N_{\rm th}$ within a relevant transmission bandwidth. (ii) OM phonon routers ensure a directed emission and complete reabsorption of individual phonons at the second node. To achieve the reverse state transfer the directionality of the routers can be reversed as described in more detail in \sref{sec:PhononRouter}.}
\label{fig:StateTransferSetup}
\end{center}
\end{figure}

The phonon network described above is formally equivalent to optical quantum networks and can in principle be used in a similar way to transfer quantum states from one node to another \cite{Cirac1997,Ritter2012}. However, in contrast to high frequency optical fields,  the thermal population $N_{\rm th}= (\exp(\hbar \omega_m/k_B T) -1)^{-1}$ of the phonon modes is usually not negligible. For phonon frequencies in the MHz to GHz regime and temperatures $T\sim 0.1-1$ K the number of thermal noise phonons in the waveguide will  exceed the quantum signal encoded in a single phonon. 
In addition, as shown in \fref{fig:PhononNetwork}c, phonons emitted from a single side-coupled node will naturally propagate along two directions, i.e. $\gamma_L=\gamma_R$. In larger networks or 2D arrangements the inability to route propagating phonons severely limits an efficient implementation of quantum communication protocols. 

In the remainder of the paper we consider an extended setup as shown in \fref{fig:StateTransferSetup} and describe how the integration of additional OM control elements can be used to overcome these fundamental limitations of phononic networks. The two key ingredients in this setup are:
\begin{enumerate}
\item An OM noise filter, which is used to suppress thermal noise within the relevant transmission bandwidth.
\item Coherent OM phonon circulators or phonon routers, which allow for directed propagation and switchable routing of phonons through the network.
\end{enumerate}

\section{Optomechanical noise filters}\label{sec:NoiseEater}

\begin{figure}[t]
\begin{center}
\includegraphics[width=0.8\linewidth]{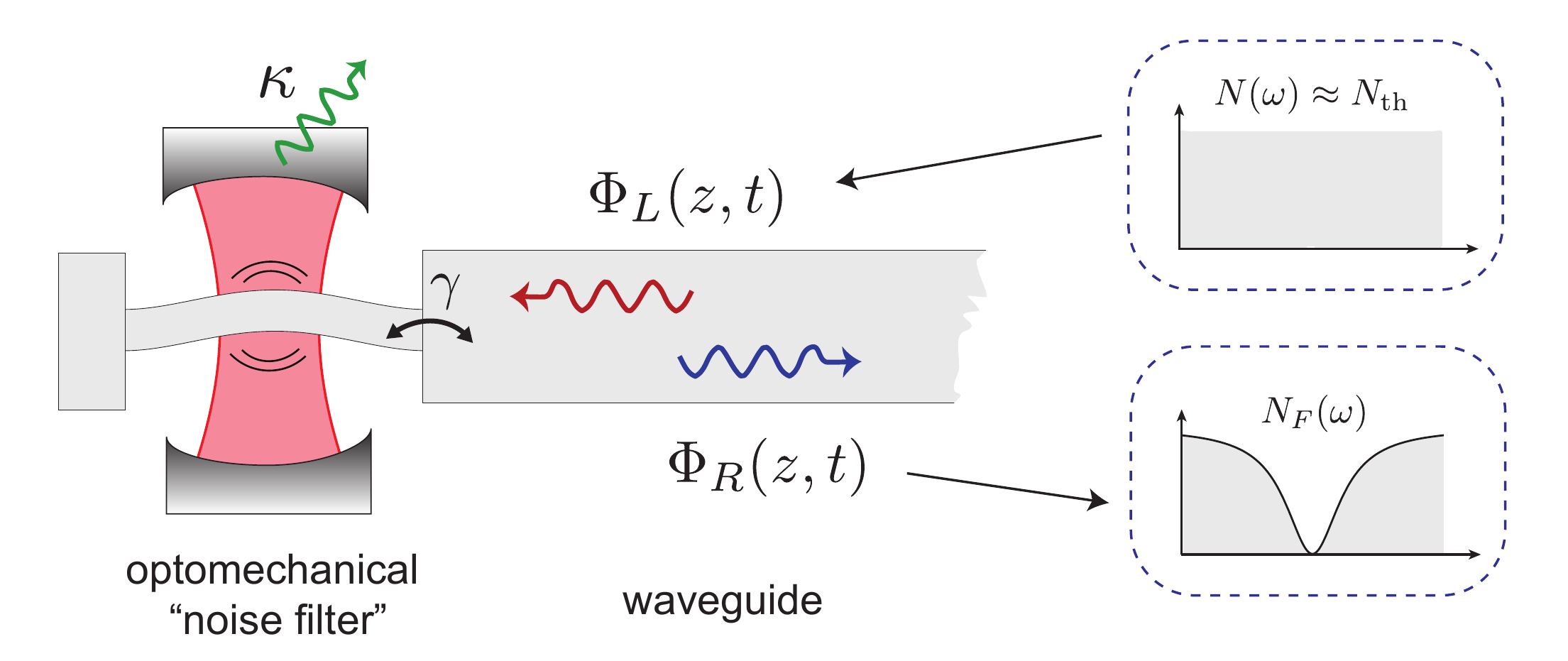}
\caption{Setup for an OM noise filter (OM continuous mode cooling). A mechanical resonator is coupled to a continuous phonon waveguide ($\gamma$) and laser cooled via an optical cavity mode. Within a bandwidth $\sim \gamma$ around the mechanical frequency, the incident thermal noise can be efficiently suppressed, creating a dip in the filtered noise spectrum $N_F(\omega)$ of the reflected waveguide modes.}
\label{fig:NoiseEater}
\end{center}
\end{figure}

In this section we describe the application of OM laser cooling techniques to suppress thermal noise in an extended phononic quantum channel. Obviously, ground-state cooling of the whole network becomes inefficient as the system size increases, but it is also unnecessary, since usually only a few modes within a small bandwidth are used for transmitting quantum states. In the following we consider a scenario as shown in \fref{fig:NoiseEater}, where a single laser-cooled mechanical resonator provides a `cold sink', and within a certain frequency range suppresses the thermal noise of the reflected modes of a continuous waveguide (or the transmitted modes in the case of a side-coupled phonon cavity). A similar  configuration has been previously proposed for realizing a traveling-wave photon to phonon converter, where under `impedance-matched' conditions an incoming photon is converted into a traveling phonon and vice versa~\cite{Safavi-NaeiniNJP2011}. In this sense cooling of the waveguide can be interpreted as a mapping of the optical vacuum onto the phonon channel, while in turn mechanical noise is upconverted into optical photons leaving the system.    

\subsection{Single mode cooling} 

Before addressing the suppression of thermal noise in extended phonon waveguides, let us first briefly review with the standard 
scenario for OM cooling \cite{Marquardt2007,WilsonRae2007}, where the frequency $\omega_c$ of an optical cavity mode is modulated by the displacement of a single mechanical mode. The optical cavity is driven by a coherent laser field of frequency $\omega_d=\omega_c+\delta $ and in a frame rotating with $\omega_d$ the Hamiltonian for the OMS is given by 
\begin{equation}
\label{Hfull}
H_{\rm om}= - \delta a^{\dag}a+\omega_{m}b^{\dag}b+ga^{\dag}a(b+b^{\dag})+i\mathcal{E}(a^{\dag}-a)\;.
\end{equation}
Here the bosonic operators $a$ and $b$ represent the optical and the mechanical modes, respectively, $\omega_m$ is the mechanical oscillation frequency, $g$ the OM coupling constant and $\mathcal{E}$ the strength of the external driving field. Terms rotating with the optical frequency have been neglected by a rotating wave approximation (RWA). The OM interaction, as described by the third term in equation \eref{Hfull}, derives from a linear dependence of the optical resonance frequency on the position quadrature of the mechanical mode \cite{Law1995,Kippenberg2007}. Including dissipation through cavity decay and intrinsic mechanical losses, the dynamics of the OMS is described by the QLEs
\begin{eqnarray}
\dot{a}&=&(i\delta-\kappa)a-ig a (b+b^{\dag})+\mathcal{E} -\sqrt{2\kappa\,}a_{\rm in},\label{eq:QLE_Single_a}\\
\dot{b}&=&(-i\omega_{m}-\gamma_0/2)b-ig a^{\dag}a-\sqrt{\gamma_0}\, b_{0,{\rm in}},\label{eq:QLE_Single_b}
\end{eqnarray}
where $\kappa$ is the optical field decay rate and $\gamma_0=\omega_m/Q_0$ is the mechanical damping rate for an intrinsic mechanical quality factor $Q_0$. The $\delta$-correlated operators $a_{\rm in}$ and $b_{0,\rm in}$ represent the vacuum input noise of the optical field and the thermal mechanical noise respectively. Within the relevant frequency range the latter is characterized by a non-vanishing equilibrium occupation number $\langle b_{0,\rm in}^{\dag}(t) b_{0,\rm in}(t')\rangle=N_{\rm th} \delta(t-t')$. 

In the limit of strong driving the external field displaces both the optical and the mechanical modes by a large classical amplitude $\alpha=\langle a\rangle =\mathcal{E}/(\kappa-i\delta(\alpha))$ and $\beta=\langle b\rangle =-g|\alpha|^{2}/\omega_{m}$, where $\delta(\alpha)=\delta+2g^2|\alpha|^{2}/\omega_{m}$. For $|\alpha|\gg1$, we can make a unitary transformation $a\rightarrow\alpha+a$ and $b\rightarrow\beta+b$ and linearize the OM coupling around the classical mean values,
\begin{equation}
\label{Hlinear}
H_{\rm om}\simeq -\delta a^{\dag}a+\omega_{m}b^{\dag}b+g(\alpha a^{\dag}+\alpha^{\ast}a)(b+b^{\dag})\;,
\end{equation}
where we have redefined $\delta(\alpha)\rightarrow \delta$.
If the external driving field is (near) resonant with the red mechanical sideband of the optical cavity, i.e. if $\delta\simeq-\omega_{m}$, and if $\omega_m\gg \kappa, |g \alpha|$ we can make a rotating wave approximation (RWA) with respect to $\omega_m$ and obtain a beam-splitter Hamiltonian
\begin{equation}
\label{eq:Beamsplitter}
H_{\rm om}\simeq H_{\rm bs}= (\omega_{m}+\delta)b^{\dag}b+g(\alpha b a^{\dag}+\alpha^{\ast}b^{\dag}a)\;,
\end{equation}
which describes a (near) resonant conversion of phonons into photons (and vice versa). Combined with the decay from the optical cavity, it allows for cooling: incident (noise) phonons are up-converted to photons and decay from the optical cavity.

The QLEs for the linearized OMS can be conveniently solved in the Fourier domain as outlined in more detail in \ref{app:OMCooling}. In the relevant weak coupling and sideband resolved regime ($|g\alpha|< \kappa\ll \omega_m=-\delta$)  we then obtain the standard result for the mechanical fluctuation spectrum, 
\begin{equation}\label{eq:SingleModeCoolingResult}
\langle b^\dag(\omega) b(\omega') \rangle\simeq  \left[  \frac{(\gamma_0+\gamma_{\rm op}) \bar N }{(\omega-\omega_m)^2+(\gamma_0+\gamma_{\rm op})^2/4}\right]\delta(\omega-\omega')\;,
\end{equation}
where $\gamma_{\rm op}=2 |g\alpha|^2/\kappa$ is the optically induced mechanical damping rate and $\bar N=N_{\rm th} \gamma_0/(\gamma_0+\gamma_{\rm op})+ \kappa^2/(4\omega_m^2)$ the reduced steady state resonator occupation number. In the following we are interested in OMS where both $\gamma_{\rm op}\gg \gamma_0$ and single-mode ground-state cooling $\bar N < 1$ can be achieved. 

\subsection{Optomechanical cooling in a multimode system} \label{sec:MultiMode}
\begin{figure}[t]
\begin{center}
\includegraphics[width=0.9\linewidth]{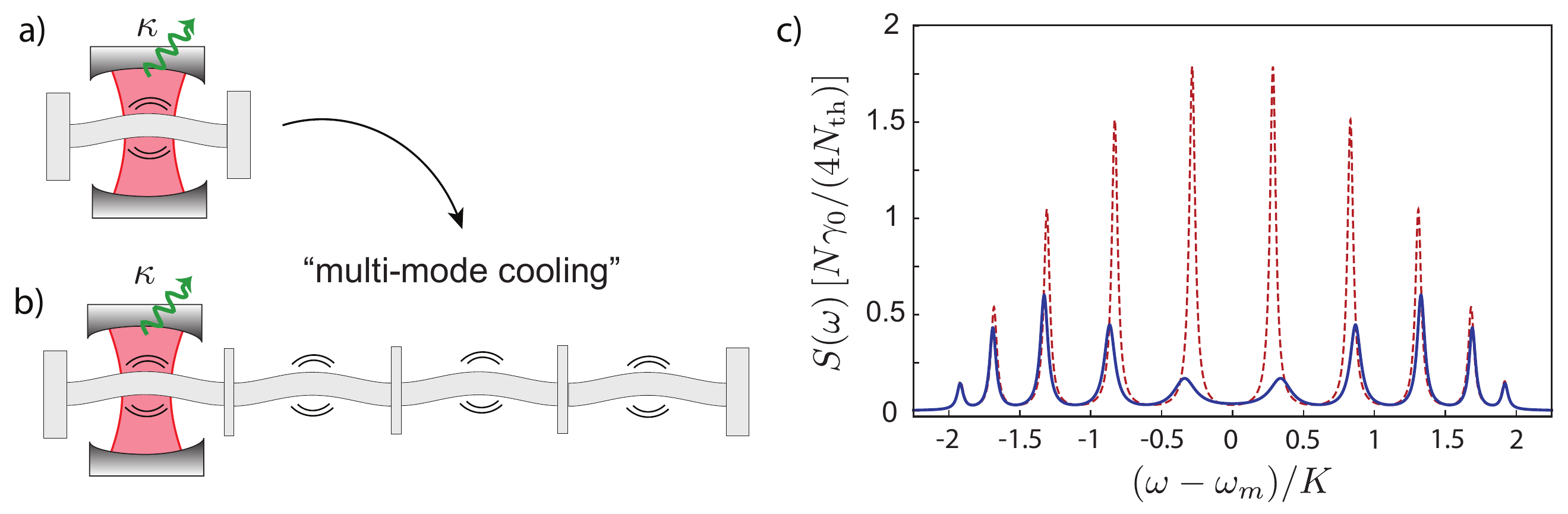}
\caption{a) Setup for OM cooling of a single mechanical mode and b)
its generalization to multiple coupled resonators. c) The steady state
fluctuation spectrum $S(\omega)$ as defined in the text is plotted for
the $b_{10}$ mode in an array of $N=10$ resonators, assuming that the
$b_1$ mode is OM cooled. The dashed line shows the result in the
absence of OM cooling, i.e. $g|\alpha|=0$, while the solid line
corresponds to $g|\alpha|/K=0.5$. The other parameters for this plot
are $\delta=-\omega_{m}$, $\gamma_{0}/K=0.05$ and $\kappa/K=0.5$.}
\label{fig:Multimode}
\end{center}
\end{figure}

The single mode cooling scheme can be generalized to cool mechanical networks consisting of a few coupled resonators only one of which is OM cooled, as shown in \fref{fig:Multimode}b. In this case the linearized OM Hamiltonian is given by 
\begin{equation}\label{eq:HMultiMode}
H_{\rm om}\simeq
 -\delta a^{\dag}a  + \sum_{j=1}^N \omega_{m}b_j^{\dag}b_j- \sum_{\langle i,j\rangle }  K_{i,j} \left(b_i b_{j}^\dag + b_i^\dag b_{j}\right) +g\alpha\left(a^{\dag} b_1+ ab_1^\dag\right)\;, 
\end{equation}
where the $K_{i,j}$ denote the nearest neighbor couplings and we have already performed a RWA assuming that $K_{i,j},g|\alpha|,\kappa\ll  \omega_m$.
The resulting QLEs can be written  as  
\begin{eqnarray}
\dot{a}&=&(i\delta-\kappa)a-i\alpha g b_1-\sqrt{2\kappa}a_{\rm in} \label{eq:QLE_Multi_a}\\
\dot{b}_1&=&-(i\omega_{m}+\gamma_0 / 2)b_1+iK_{1,2} b_2 -ig\alpha^{\ast}a -\sqrt{\gamma_0}b_{0,{\rm in}}^{(1)}\\
\dot{b}_{1<j<N}&=&-(i\omega_{m}+\gamma_0 /2)b_j+i(K_{j-1,j} b_{j-1} + K_{j,j+1}b_{j+1}) -\sqrt{\gamma_0}b_{0,{\rm in}}^{(j)} \\
\dot{b}_N&=&-(i\omega_{m}+\gamma_0 / 2)b_N+i K_{N-1,N} b_{N-1} -\sqrt{\gamma_0}b_{0,{\rm in}}^{(N)} \label{eq:QLE_Multi_bN}\;, 
\end{eqnarray}
where the $b_{0,{\rm in}}^{(j)}$ are mutually uncorrelated noise operators associated with intrinsic mechanical damping of each mode. 

In \fref{fig:Multimode}c we plot the correlation spectrum $\langle b_N^\dag(\omega)b_N(\omega')\rangle=S(\omega)\delta(\omega-\omega')$ for a chain of $N=10$ resonators with $K_{i,i+1}=K$ assuming that the $b_1$ mode is optically cooled. For reference the dashed line shows the undamped case $g\alpha=0$. In this case  the 
heights of the peaks at frequencies $\omega_n=\omega_m-2K\cos(n \pi/(N+1))$ are given by $S(\omega_n)=(4N_{\rm th}/\gamma_0) |c_n(N)|^2$, where $c_n(j)=\sqrt{2/(N+1)}\sin( n j\pi/(N+1))$ are the normalized mode distributions. When the OM coupling is turned on the peaks are broadened and their height is reduced, which corresponds to an overall cooling of the individual eigenmodes. However, we see that cooling occurs in a highly non-uniform way and only the few modes close to the cavity resonance are cooled efficiently. While the details depend on the ratios between $K$, $g|\alpha|$ and $\kappa$, we find that this behavior is quite generic and a precursor to the features we identify in the following for the continuous waveguide limit. 

\subsection{Optomechanical cooling of a mode continuum} \label{sec:NoiseFilter}
Starting from the multi-mode setting shown in \fref{fig:Multimode}b, let us now address the limit $N\rightarrow \infty$ of a continuous mode waveguide as depicted in \fref{fig:NoiseEater} by retaining the OM coupling to the first mode $b\equiv b_1$, but describing the other phonon modes $b_{j>1}$ in terms of continuous left- and right-propagating fields $\Phi_{L,R}(z)$. If  $K_{1,2}\ll K_{i,i+1}\equiv K$, we can adapt the input-output formalism introduced in \sref{sec:PhononNetworks}  to model the coupling between the localized mode $b$ and the rest of the waveguide in terms of 
 the scattering relation 
\begin{equation}
b_{\rm out}(t)= b_{\rm in}(t) + \sqrt{\gamma} b(t).
\end{equation} 
Here $b_{\rm in} (t)= \Phi_L(z\rightarrow0,t)$ and $b_{\rm out}(t)= \Phi_R(z\rightarrow 0,t)$ are the incident and reflected waveguide fields and $\gamma\approx K_{1,2}^2/K$
is the characteristic phonon decay rate into the waveguide. Ignoring other, intrinsic loss channels for the moment, we obtain   
\begin{eqnarray}
\dot{a}&=&(i\delta-\kappa)a-i\alpha g (b+b^{\dag})-\sqrt{2\kappa}a_{\rm in}\label{HLeq_a}\\
\dot{b}&=&-(i\omega_{m}+\gamma/2)b-ig(\alpha a^{\dag}+\alpha^{\ast}a)-\sqrt{\gamma}b_{{\rm in}} \label{HLeq_b}\;,
\end{eqnarray}
for the linearized OM dynamics of the local mode and  we see that the problem of cooling a continuous waveguide is formally identical to the single-mode cooling considered above. However, instead of looking at the occupation of the local mode $b$ we are now interested in the spectrum of the reflected waveguide field $b_{\rm out}(t)$, given a thermal incident field $\langle b^\dag_{\rm in}(t)b_{\rm in}(t')\rangle= N_{\rm th} \delta(t-t')$. To do so, we solve the QLEs in Fourier space and write the result as
\begin{equation}
\label{inoutS}
\vec A_{\rm out}(\omega)=\mathcal{S}(\omega)\vec A_{\rm in}(\omega)\;,
\end{equation}
where $\vec{A}_{j}=(a_{j}(\omega),a^{\dag}_{j}(-\omega),b_{j}(\omega),b^{\dag}_{j}(-\omega))^{\rm T}$  with $j=$ ``in'', ``out''. The matrix $\mathcal{S}(\omega)$ is a $4\times 4$ scattering matrix and a more detailed derivation of equation \eref{inoutS} is given in \ref{app:OMCooling}.  For given input noise correlations $C_{\rm in}(\omega,\omega^\prime)=\langle\vec A_{\rm in}(\omega)\otimes \vec A_{\rm in}^{\rm T}(\omega')\rangle$ we obtain the output correlation matrix  $C_{\rm out}(\omega,\omega^\prime)=\langle\vec A_{\rm out}(\omega)\otimes \vec A_{\rm out}^{\rm T}(\omega')\rangle=\mathcal{S}(\omega)C_{\rm in}(\omega,\omega')\mathcal{S}^{\rm T}(\omega')$.

\begin{figure}[t]
\begin{center}
\includegraphics[width=0.8\linewidth]{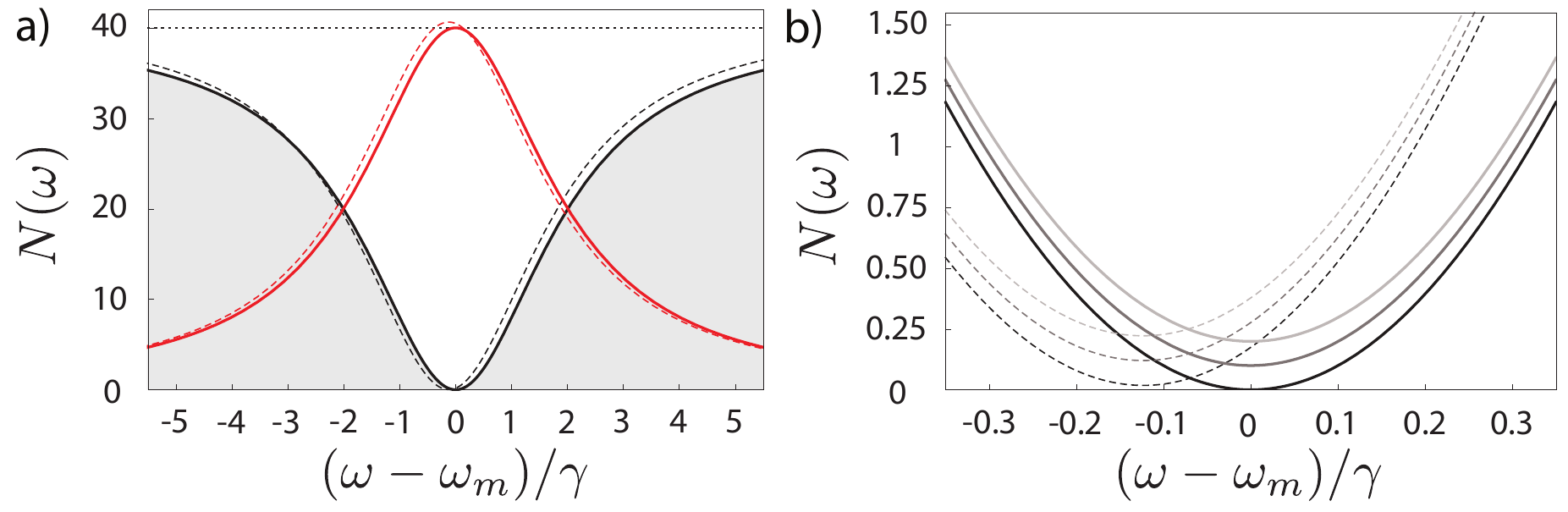}
\caption{Detailed results for OM noise filtering. a) Filtered
phononic noise spectrum $N_{\rm F}(\omega)$ (black lines) and the
corresponding photonic output spectrum (red lines), with the
beam-splitter Hamiltonian \eref{eq:Beamsplitter} (solid lines) and
with the full linearized Hamiltonian \eref{Hlinear} (dashed lines).
The parameters are $\omega_{m}/\gamma=1200$, $\gamma_{0}=0$,
$\kappa/\gamma=300$,
$g|\alpha|=\sqrt{(\gamma+\gamma_0)\kappa/2}$ and $N_{\rm th}=40$. b)
Dip of the filtered noise spectrum $N_{\rm F}(\omega)$ for different
values of the intrinsic mechanical damping rate $\gamma_{0}/\gamma=0,\ 0.0025,\ 0.005$.}
\label{fig:NoiseEaterResults}
\end{center}
\end{figure}

\subsubsection{Optomechanical noise filters.}
The quantity that we are interested in is the filtered noise spectrum $N_F(\omega)$ of the reflected field, which is defined by
\begin{equation}
\langle b_{\rm out}^{\dag}(\omega) b_{\rm out}(\omega')\rangle=N_{\rm F}(\omega) \delta(\omega-\omega')\;.
\end{equation} 
\Fref{fig:NoiseEaterResults}a shows the typical frequency dependence of $N_F(\omega)$. As expected, we see a strong suppression of thermal noise around the mechanical frequency $\omega_m$.  For $\delta=-\omega_m$ and under the condition $|g\alpha|< \kappa \ll \omega_m$, the dynamics of the OMS is well approximated by the beam-splitter Hamiltonian \eref{eq:Beamsplitter} and we obtain  
\begin{eqnarray}
N_{F} (\omega)\simeq N_{\rm th}\left(1\!-\! \frac{4 \kappa^2 \gamma_{\rm op} \gamma}{\kappa^2(\gamma_{\rm op} \!+\! \gamma)^2 +(\gamma\!-\!2\kappa)^{2}(\omega-\omega_{m})^2+4(\omega\!-\!\omega_{m})^{4}}\right)\;. \label{eq:NFideal}
\end{eqnarray}
We see that, under these approximations, the `impedance matching' condition $\gamma_{\rm op}=\gamma$ \cite{Safavi-NaeiniNJP2011} results in a complete cancellation $N_F(\omega= \omega_m)=0$ of the reflected noise on resonance. In this case the optical decay rate matches the mechanical waveguide coupling and the OMS acts effectively as a double-sided phonon cavity with the thermal mechanical bath on one side and the optical $T=0$ bath on the other side. 

Under realistic conditions the noise cancellation described by equation \eref{eq:NFideal} will not be perfect and to account for various imperfections in the system    
we assume in our discussion below a spectrum of the form 
\begin{equation}\label{eq:NF}
N_{F} (\omega)=  N_{\rm th}-  (N_{\rm th}-N_0)  \frac{ \tilde \gamma^2}{(\omega-\tilde \omega_m)^2 + \tilde \gamma^2},
\end{equation}
where $\tilde \gamma$ and $\tilde \omega_m$ now include  small corrections of the width and the center of the dip and $N_0$  is a non-vanishing noise floor. Intrinsic limitations that lead to a finite $N_0$ are already contained in the full OM interaction itself, where for any finite $\omega_m$ off-resonant Stokes-scattering terms in the linearized Hamiltonian \eref{Hlinear} induce corrections to the ideal beam-splitter interaction. However,  as shown in \fref{fig:NoiseEaterResults}b, by accounting for these effects up to second order in  $1/\omega_{m}$ we only obtain a small shift of the spectral dip $\tilde \omega_m\simeq\omega_m-g^2 |\alpha|^2/(2\omega_{m})^2$ and an offset $N_0\simeq |g\alpha|^2/(4\omega_m^2)$, which is negligible under the weak coupling conditions mentioned above. A more crucial limitation  for the OM noise filter arises from intrinsic mechanical losses $\gamma_0$ of the localized mechanical mode, which can be accounted for by introducing an additional decay rate $\gamma_0$ and associated noise operator $b_{\rm 0,in}(t)$ in the QLE \eref{HLeq_b} (see \ref{app:OMCooling}). This leads to a finite $N_{0}\simeq 4N_{\rm th}\gamma\gamma_{0}/(\gamma_{\rm op}+\gamma+\gamma_{0})^2$. This means that, while for the optimized case with $\gamma=\gamma_0+\gamma_{\rm op}$ the local mode $b$ can be highly excited, also the condition $\gamma_0 N_{\rm th}/\gamma_{\rm op}\ll 1$ must be fulfilled to achieve good noise filtering.

\subsubsection{Propagation losses and scattering.} The spectrum $N_F(\omega)$ determines the mechanical noise of the outgoing displacement field $b_{\rm out}(t)\equiv\Phi_R(z\rightarrow0,t)$ immediately after the OM device and  for an ideal phonon waveguide this noise spectrum will be the same at all positions $z>0$.  However, in a realistic setting, scattering losses in the waveguide and the associated noise lead to a rethermalization of the noise spectrum  and  $N_F(\omega)\rightarrow N_F(\omega, z)$. For an approximately linear dispersion relation the propagation of the outgoing field can be modeled by
\begin{equation}
\left(\frac{\partial}{\partial t}+ c \frac{\partial}{\partial z}\right) \Phi_R(z,t) = -\left(i\tilde \omega_0 +\frac{c}{2l_\gamma} \right)  \Phi_R(z,t) -\frac{c}{ \sqrt{l_\gamma}} \Phi_{\rm th}(z,t)\;,
\end{equation}
where $l_\gamma$ is the phonon mean free path in the waveguide  and $\Phi_{\rm th}(z,t)$ is a thermal noise field with  $\langle   \Phi^\dag_{\rm th}(z,t) \Phi_{\rm th}(z',t')\rangle =N_{\rm th} \delta(z-z')\delta(t-t')$. In \ref{app:ResonatorArray:Losses} we outline a microscopic derivation of  this result for a coupled mechanical resonator array, where $l_\gamma=c/\gamma_0$ can be connected to the intrinsic damping rate $\gamma_0$ of the individual resonators in the array.  For $z>0$ and in a frame rotating with  $ \tilde \omega_0$ we obtain
\begin{eqnarray}
\Phi_R(z,t) = e^{-\frac{z}{2l_\gamma} }   \Phi_R\left(0,t-\frac{z}{c}\right) - \frac{1}{\sqrt{ l_{\gamma}}} \int_0^z dz' \,  e^{-\frac{(z\!-\!z')}{2l_\gamma}  }\Phi_{\rm th}\left(z',t-\frac{(z\!-\!z')}{c}\right)\;,\nonumber\\
\end{eqnarray}
and from this result we can derive the full noise spectrum along the waveguide
\begin{equation}
N_F(\omega,z)= e^{-\frac{z}{l_\gamma} }  N_F(\omega,0) + N_{\rm th}\left(1- e^{-\frac{z}{l_\gamma}}\right).
\end{equation}
This result describes the rethermalization of the noise dip on a length scale given by the phonon mean free path. For the relevant distances $z\ll l_\gamma$ the noise spectrum can still be described by the Lorentzian shape given in equation \eref{eq:NF}, but the noise floor $N_0(z)\simeq N_0+ (z/l_\gamma)N_{\rm th}$ increases linearly with the distance to the OM noise filter.  Note that similar quantitative conclusions follow from the standard approach of treating propagation losses in terms of a series of beam splitters \cite{Safavi-NaeiniNJP2011,QuantumNoise}, but the present analysis also provides a direct connection to the underlying microscopic phonon scattering mechanism.

To the extend that the scatterers are linear, phonon backscattering gives rise to small overall losses, which are on equal footing with the thermalization discussed above.

\section{Quantum state transfer in a thermal phonon network} \label{sec:StateTransfer}
We now return to our original goal of implementing a quantum state
transfer protocol between two qubits via an extended and thermally
occupied phononic channel. For a simplified discussion we consider in the following the state transfer between two nodes assuming a unidirectional
coupling, where $\gamma\equiv\gamma_R$, $\gamma_L=0$ and $b_{i,{\rm in}/{\rm out}}\equiv
b^R_{i,{\rm in}/{\rm out}}$ as shown in \fref{fig:StateTransferSetup}.  In \sref{sec:PhononRouter} below, we describe how this condition can be realized using phonon routers.

\subsection{A tunable qubit network}
As a first step we describe the implementation of an effective qubit network with tunable qubit-waveguide couplings $\Gamma_{1,2}(t)$ by eliminating the dynamics of 
the local phonon modes. We start from the full set of QLEs, which, for each node $j=1,2$, derive from the Hamiltonian in \eref{eq:Hnode} and are given by
\begin{eqnarray}\label{eq:QLE1}
\dot{\sigma}_-^j&=&-i\Delta_q(t) \sigma_-^j + i\lambda_j(t)\sigma_z^j
b_j \;,\\
\dot{\sigma}_z^j&=&-2i\lambda_j(t)( \sigma_+^j b_j - b_j^\dag
\sigma_-^j) \label{eq:QLE2}\;,\\ 
\dot{b}_j&=&-i\omega_m b_j-\frac{\gamma}{2}  b_j -i\lambda_j(t)\sigma^j_-
-\sqrt{\gamma}b_{j,{\rm in}}\;.\label{eq:QLE3}
\end{eqnarray}
The input-output relations are 
\begin{equation}
b_{j, {\rm out}}(t)=  b_{j,{\rm in}}(t) + \sqrt{\gamma} b_j(t)\qquad\mathrm{and}\qquad  b_{2,{\rm 
in}}(t)=  b_{1,{\rm out}}(t-\tau_{12})\;,
\end{equation}
where $b_{\rm 1,in}(t)\equiv b_{\rm in}(t)$ describes the incident waveguide field before interacting with the first node.  Hereafter, we absorb the retardation time $\tau_{12}$ by redefining time-shifted operators and control fields for the second node~\cite{QuantumNoise} and for simplicity we focus  on the resonant case in which $\Delta^j_q(t)=\omega_m$. 

We are interested in the regime in which the decay $\gamma$ of the the local phonon modes into the waveguide is fast compared to the coupling $\lambda(t)$. This allows us to adiabatically eliminate the phonon modes and to derive an effective description in terms of the qubit degrees of freedom only. In the frame rotating with $\omega_m$ and to lowest order in $\lambda$, \eref{eq:QLE3} can be solved to give
\begin{equation}
    b_j(t)\simeq  -\frac{2i\lambda_j(t)}{\gamma} \sigma_-^j(t)-
\sqrt{\gamma} \int_{-\infty}^t ds \,
e^{-\frac{\gamma}{2}(t-s)} \, e^{i\omega_m s} b_{j,{\rm in}}(s)\;.
   \end{equation}
After reinserting this expression into equations \eref{eq:QLE1} and \eref{eq:QLE2} we obtain 
  \begin{eqnarray}
\label{eq:qubitQLE1}
\dot{\sigma}_-^j&=&- \frac{\Gamma_j(t)}{2}
\sigma_-^j -\sqrt{\Gamma_j(t)} \sigma_z^j B_{j,{\rm in}}\\
\label{eq:qubitQLE2}
\dot{\sigma}_z^j&=&-\Gamma_j(t)\left(\mathbbm{1} +\sigma_z^j\right) +
2\sqrt{\Gamma_j(t)}\left( \sigma_+^jB_{j,{\rm in}} +B_{j,{\rm in}}^\dag
\sigma_-^j\right)\;.
\end{eqnarray}
Here $\Gamma_j(t)= 4\lambda_j^2(t)/\gamma$ are tunable qubit decay rates and 
\begin{equation}
B_{j,{\rm in}}(t)=\frac{i\gamma}{2} \int_{-\infty}^t ds \,
e^{-\frac{\gamma}{2}(t-s)} \, e^{i\omega_m s}b_{j,{\rm in}}(s)\;,
\end{equation}
the associated effective noise operators which on the timescale
$\gamma^{-1}$ obey 
$[B_{j,{\rm in}}(t),B_{j,{\rm in}}^\dag(t')]\approx  \delta(t-t')$. Provided that $\lambda(t)$ and $\sigma_-^1(t)$ also vary slowly on this time scale, we find that 
\begin{equation}\label{eq:InOutQubit}
B_{2,{\rm in}}(t)\simeq -  B_{1,{\rm in}}(t) + \sqrt{\Gamma_1(t)} \sigma_-^1(t)\;.
\end{equation}
Thus we obtain a new set of effective quantum network equations for the qubit operators with tunable decay rates $\Gamma_j(t)$. We emphasize that while in the following this effective description allows a simplified discussion of the state transfer protocol, it is not necessary to consider the regime $\lambda\ll \gamma$ and a perfect state transfer can also be achieved using the full model \cite{Cirac1997}.

\subsection{Quantum state transfer through a phonon channel}
Our goal is to implement a quantum state transfer between the two qubits, i.e.
\begin{equation}
\label{eq:InitialState}
|\psi_0\rangle=(\alpha|0\rangle_1+\beta |1\rangle_1)|0\rangle_2 \rightarrow |0\rangle_1(\alpha |0\rangle_2+ \beta|1\rangle_2)\;,
\end{equation}
which is achieved via coherent emission and reabsorption of a single phonon in the waveguide. A solution to this problem  has been first described  for optical networks \cite{Cirac1997}, where it has been shown that by identifying appropriate pulses  $\Gamma_{j}(t)$, a perfect quantum state transfer can be implemented. Let us briefly summarize the main idea  behind this scheme, for the moment assuming zero temperature. In this case the waveguide is initially in the vacuum state $|{\rm vac}\rangle$ and the dynamics of the whole system can be restricted to the  zero- and one-excitation subspace. Then, for an initial two qubit state $|\psi_0\rangle$, we can define the amplitudes $v_j(t)=\langle {\rm vac}|\langle0_1, 0_2| \sigma_-^j(t)|\psi_0\rangle|{\rm vac}\rangle$. From the QLEs (\ref{eq:qubitQLE1}) and (\ref{eq:qubitQLE2}) and the input-output relation (\ref{eq:InOutQubit}), it follows that these amplitudes evolve according to
\begin{eqnarray}
\dot v_1(t)&=&-\frac{\Gamma_{1}(t)}{2}v_1(t), \label{eq:v1dot}\\
\dot v_2(t)&=&-\frac{\Gamma_{2}(t)}{2}v_1(t)-\sqrt{\Gamma_{1}(t)\Gamma_2(t)}v_1(t),\label{eq:v2dot}
\end{eqnarray}
and for initial amplitudes $v_j(t_0)$, the general solutions is given by 
\begin{eqnarray}
v_1(t)&=&\mathcal{G}_{1}(t,t_{0})v_1(t_0)\;,\\
v_2(t)&=&\mathcal{G}_{2}(t,t_{0})v_2(t_0)+\mathcal{T}(t,t_{0})v_1(t_0)\;.
\end{eqnarray}
Here $\mathcal{G}_{j}(t,t_{0})=e^{-\int_{t_{0}}^{t}ds\;\Gamma_{j}(s)/2}$ and 
\begin{equation}
\mathcal{T}(t,t_{0})=-\int_{t_{0}}^{t}dt'\; \mathcal{G}_{2}(t,t')\sqrt{\Gamma_{1}(t')\Gamma_{2}(t')}\mathcal{G}_{1}(t',t_{0})\;,
\end{equation}
is the state-transfer amplitude. For the initial state given in equation \eref{eq:InitialState}, $v_{2}(t_{0})=0$ and therefore a perfect transfer is achieved if at some final time $t_{f}$ the amplitudes approach
\begin{equation}
\mathcal{G}_1(t_f,t_0)\rightarrow 0\qquad\mathrm{and}\qquad |\mathcal{T}(t_f,t_0)|\rightarrow 1\;.
\end{equation}
While the first condition is easily fulfilled for sufficiently long, but otherwise arbitrary pulses, satisfying the second one requires control over their shape. Perfect state transfer is only possible if the total number of excitations is conserved, i.e., if no population gets lost from the one-excitation subspace. This means that the norm $\mathcal{G}_1^2(t,t_0)+\mathcal{T}^2(t,t_0)=1$ for all times, which after taking the time-derivative of this condition yields
\begin{equation}
\label{darkstate}
\sqrt{\Gamma_{1}(t)}v_1(t)+\sqrt{\Gamma_{2}(t)}v_2(t)=0\;.
\end{equation}
This result also follows from the requirement that the total scattered field after the second node vanishes at all times, i.e. $B_{2,{\rm out}}(t)|\psi_0\rangle|{\rm vac}\rangle=0$. Therefore equation \eref{darkstate} is often referred to as the {\it dark-state condition}.

A set of pulses $\Gamma_1(t)$ and $\Gamma_2(t)$ that realize a perfect state transfer can always be found numerically by choosing $\Gamma_1(t)$ such that $\mathcal{G}_{1}(t_{f},t_{0})\rightarrow 0$ and then solve equations \eref{eq:v1dot} and \eref{eq:v2dot} iteratively, choosing  $\Gamma_{2}(t)$ at each time such that the dark-state condition \eref{darkstate} is fulfilled. Further, simple analytical expressions for pulses that realize a perfect state transfer may be obtained from a time-inversion argument \cite{Cirac1997}. Without loss of generality, we can assume that $[t_{f},t_{0}]=[-\tau_p/2,\tau_p/2]$, where $\tau_p$ is the pulse length. A control pulse $\Gamma_{1}(t)$ for the first qubit gives rise to a wave packet $a(t)=\sqrt{\Gamma_{1}(t)}\mathcal{G}_{1}(t,t_{0})v_{1}(t_{0})$ in the waveguide. For reasons of time-inversion symmetry, the inverted wave packet $a(-t)$ is fully absorbed if the reverse pulse $\Gamma_{1}(-t)$ is applied. This implies that, in the special case of a time-symmetric wave packet $a(t)=a(-t)$, the wave packet generated at the first node will be fully absorbed if we choose $\Gamma_{2}(t)=\Gamma_{1}(-t)$. Describing the symmetric wave packet by the differential equation $\dot{a}(t)=g(t)a(t)$, where $g(t)$ is anti-symmetric $g(t)=-g(-t)$, one can derive the following differential equation for the pulse-shape $\Gamma_{1}(t)$  from \eref{eq:v1dot} \cite{Stannigel2011}:
\begin{equation}
\dot{\Gamma}_{1}=\Gamma_{1}^{2}+2g(t)\Gamma_{1}\;.
\end{equation}
For the simplest choice of $g(t)$, i.e. $g(t)={\rm sign}(t)(\Gamma_{\rm max}/2)$, where $\Gamma_{\rm max}$ is the maximal decay rate, the solution reads
\begin{equation}\label{eq:GammaPulse}
\Gamma_{1}(t)=\left\{\begin{array}{ccc}\left(\frac{e^{-\Gamma_{\rm max}t}}{2-e^{-\Gamma_{\rm max}t}}\right)\Gamma_{\rm max}&{\rm if}&t<0\\
\Gamma_{\rm max}&{\rm if}&t\geq 0\end{array}\right.\quad\mathrm{and}\quad \Gamma_{2}(t)=\Gamma_{1}(-t)\;.
\end{equation}
We will use this specific pulse shape for our numerical examples discussed below.

\subsection{State transfer through a noisy quantum channel}
Now we consider the same state-transfer problem but in the case in which the in-field of the quantum channel is characterized by a non-vanishing noise spectrum $\langle B^\dag_{{\rm in}}(\omega) B_{{\rm in}}(\omega')\rangle =N(\omega)\delta(\omega-\omega')$, which can either be a flat thermal background $N(\omega)= N_{\rm th}$ or the filtered noise dip $N(\omega)\simeq N_F(\omega)$ as discussed in \sref{sec:NoiseFilter}.

\begin{figure}[t]
\begin{center}
\includegraphics[width=0.9\linewidth]{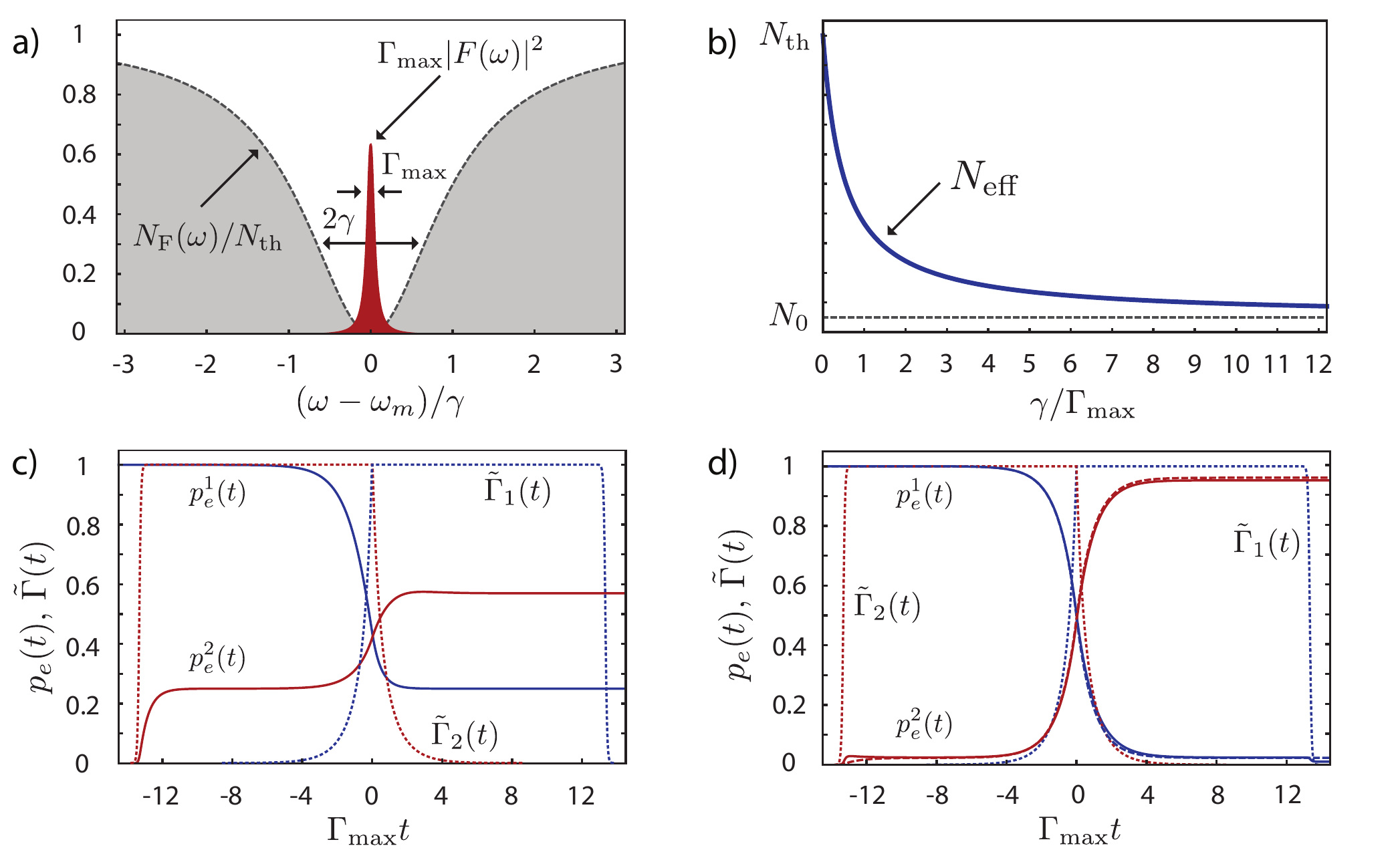}
\caption{Quantum state transfer through a noisy channel. a) Spectral
overlap of the transfer pulse $|F(\omega)|^2$  and the filtered noise
spectrum $N_F(\omega)$ for $N_{0}=0$ and $\Gamma_{\rm
max}/\gamma=0.10$. b) Effective occupation number $N_{\rm eff}$ as a
function of $\gamma/\Gamma_{\rm max}$ for $N_{0}/N_{\rm th}=0.05$. c)
Numerical simulation of a quantum state transfer in the presence of
thermal noise with $N_{\rm th}=0.5$. d) The same state transfer with
the OM noise filter switched on and $\Gamma_{\rm max}/\gamma=0.10$. In
both plots the blue and red solid lines show the excited state
populations $p_{e}^{i}(t)=\langle\sigma^i_{+}(t)\sigma^i_{-}(t)\rangle$ for the 
qubits 1 and 2, respectively. The dotted lines show the corresponding
pulse shapes $\tilde \Gamma_i(t)=\Gamma_{i}(t)/\Gamma_{\rm max}$ as defined in equation \eref{eq:GammaPulse}, but with cut-offs. In d) the dashed line shows the results obtained from a two qubit cascaded master equation with an effective thermal occupation number $N_{\rm eff}$.}
\label{fig:NoisyTransfer}
\end{center}
\end{figure}

\subsubsection{Effective thermal occupation number.} In the presence of thermal excitations in the quantum channel we must necessarily go beyond the single excitation subspace and a closed analytic solution to the state transfer problem is no longer available. To gain more intuition on the impact of noise on the state-transfer fidelity, let us first consider a single qubit.  We assume that at time $t=t_{0}$ the qubit is prepared in its groundstate and we then switch on the coupling to the waveguide $\Gamma_1(t)$, for example, using the pulse shape defined in equation (\ref{eq:GammaPulse}). In the regime where the noise amplitude is low, $N(\omega)\ll 1$, we can linearize the QLEs \eref{eq:qubitQLE1} and \eref{eq:qubitQLE2} and obtain for the final qubit excitation  $\langle \sigma_+^1 (t_f)\sigma_-^1(t_f)\rangle\simeq N_{\rm eff}\ll 1$, where
\begin{eqnarray}
\label{eq:Neff}
N_{\rm eff}=\int_{t_0}^{t_f} dt_1 dt_2  \,\sqrt{\Gamma_1(t_1)\Gamma_1(t_2)} \mathcal{G}_1(t_f,t_1) \mathcal{G}_1(t,t_2)  \langle  B^\dag_{{\rm in}}(t_1)B_{{\rm in}}(t_2)\rangle.
\end{eqnarray}
In the case of incident $\delta$-correlated thermal noise and assuming that the pulse duration is sufficiently long, i.e. $t_f-t_0\gg \Gamma_{\rm max}^{-1}$, we find $N_{\rm eff}= N_{\rm th}$. This means that during the state transfer an erroneous excitation or de-excitation probability for the qubits of $\sim N_{\rm th}$ can be expected\footnote{Of course, for larger $N_{\rm th}$ the non-linearity of the qubit must be taken into account.}.  In turn, for the filtered noise spectrum $N_F(\omega)$ defined in equation \eref{eq:NF} we obtain
\begin{eqnarray}
\label{eq:NeffOverlap}
N_{\rm eff}=\int_{-\infty}^\infty d\omega \,   |F(\omega)|^2 N_F(\omega) \simeq\frac{2\gamma N_{0}+\Gamma_{\rm max}N_{\rm th}}{2\gamma+\Gamma_{\rm max}},
\end{eqnarray}
where in general  $F(\omega)=\frac{1}{\sqrt{2\pi}}\int_{t_0}^{t_f} dt \ e^{i\omega t}\sqrt{\Gamma_1(t)}\mathcal{G}_1(t_f,t)$ and the integral in equation \eref{eq:NeffOverlap} has been evaluated for the pulse shape given in equation (\ref{eq:GammaPulse}). This shows that using the OM noise filter to clean the waveguide, the effective occupation number can be considerable reduced compared to $N_{\rm th}$, assuming that $N_0\ll 1$ and that the bandwidth of the transfer pulse fits within the noise dip. This is illustrated in \fref{fig:NoisyTransfer}, where we plot the spectral overlap between $|F(\omega)|^2$ and $N_F(\omega)$ and the resulting $N_{\rm eff}$ for different ratios $\gamma/\Gamma_{\rm max}$. 

\subsubsection{Discussion.} To verify our approximate analytic predictions we convert the cascaded QLEs into an equivalent cascaded master equation~\cite{QuantumNoise} and simulate the full state transfer numerically. Since a master equation for the two qubits can only be derived for $\delta$-correlated noise, we include the OM noise filter as a third subsystem to emulate the spectral variations of the incident noise, as outlined in \ref{app:CascadedME}. In \fref{fig:NoisyTransfer}c and \fref{fig:NoisyTransfer}d we simulate the transfer of a single qubit excitation from node 1 to node 2 and compare the case of a thermal quantum channel where $N(\omega)=N_{\rm th}$ to the case where the OM noise filter is switched on and $N(\omega)=N_F(\omega)$. We see that even for $N_{\rm th}=0.5$ thermal noise in a phonon quantum network already almost completely washes out the transferred signal, while in combination with the noise filter a high fidelity state transfer is still possible.     
In \fref{fig:NoisyTransfer}d we also compare the full numerical results with a master equation for the qubits only, assuming a $\delta$-correlated noise with an effective  occupation number $N_{\rm eff}$. We find that apart from small corrections when pulses are switched on and off, this reduced model describes the state transfer very accurately, which shows that, indeed, $N_{\rm eff}$ is the relevant parameter for a quantum channel with a filtered noise spectrum.

\begin{figure}[t]
\begin{center}
\includegraphics[width=0.85\linewidth]{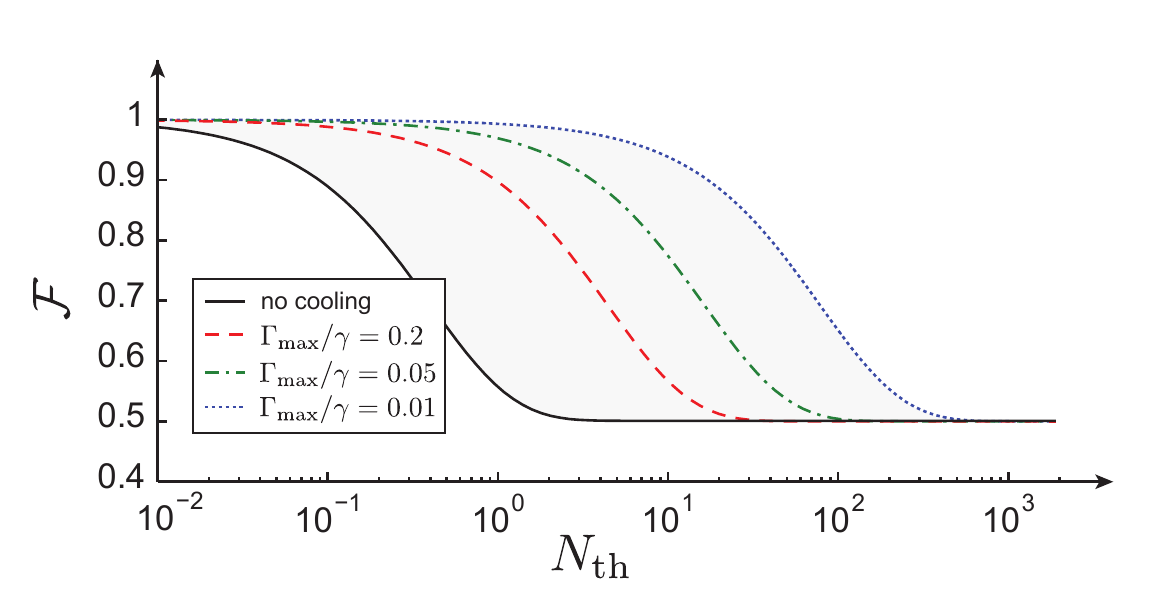}
\caption{Quantum state transfer fidelity $\mathcal{F}$ for a  superposition state $|\psi\rangle=(|0\rangle+|1\rangle) /\sqrt{2}$ as a function of the thermal occupation $N_{\rm th}$ of the waveguide and  different values of $\Gamma_{\rm max}$. For this plot we have used the pulse shape given in equation \eref{eq:GammaPulse} and assumed $\gamma_0/\gamma=1.6 \times 10^{-4}$. }
\label{fig:Fidelity}
\end{center}
\end{figure}

As a specific example, we consider a potential realization of phonon networks using  the on-chip OM structures discussed in reference \cite{Safavi-NaeiniNJP2011}. We assume local phonon modes of frequency $\omega_{m}=2\pi\times 4$ GHz and a  decay into the waveguide of $\gamma=2\pi\times 25$ MHz. For a quality factor $Q_{0}=10^{6}$ the intrinsic decoherence rate is  $\gamma_{0}=2\pi\times 4$ kHz. For these parameters we plot in \fref{fig:Fidelity} the fidelity $\mathcal{F}$ for transferring the superposition state $|\psi\rangle=(|0\rangle+|1\rangle)/\sqrt{2}$ between two nodes of a phonon network as a function of $N_{\rm th}$ and for different effective qubit decay rates $\gamma_0\ll \Gamma_{\rm max}<\gamma$. For the numerical simulations we have used the effective model with $N_{\rm eff}$ as obtained in equation \eref{eq:NeffOverlap} and with $N_{0}=\gamma_{0} N_{\rm th}/\gamma$. The fidelity is defined as the overlap between the target state and the actual state after the transfer, i.e. $\mathcal{F}={\rm Tr}\{ \rho_{\rm tar}\rho(t_{f})\}$. We see that compared to the case where no cooling is applied and the fidelity already drops significantly for $N_{\rm th}\sim 0.1$, the presence of the OM noise filter can push this bound to much larger occupation numbers. In particular, in the example considered here, this means that instead of requiring temperatures of $T\leq 100$ mK, the OM cooling scheme allows the implementation of high fidelity state transfer protocols at much more convenient liquid Helium temperatures $T=4$ K where $N_{\rm th}\sim 20$.  Note that for the assumed value of $\Gamma_{\rm max}= 10^{-2}\gamma=2\pi\times 250$ kHz, the total transfer time of $\sim1\mu$s is compatible with decoherence times that are achievable with various different solid-state qubits. A few specific examples will be discussed below in section \ref{sec:Implementations}.

\section{Phonon routers}\label{sec:PhononRouter}
In the previous sections we have studied the transfer of single excitations through a thermal phonon network assuming that two nodes are coupled in a unidirectional way. This situation is applicable only to specific setups, for example, when the two nodes are located at the two ends of a single waveguide. In general the emission of phonons into left \emph{and} right propagating modes, reflection and interference effects, or imperfect routing of phonons through multi-port junctions in a 2D setting will limit the implementation  of efficient state transfer protocols in larger networks. In optical networks, many of these problems can be overcome by using optical circulators and optical isolators for the directional routing of photons. In the context of OMS it has already been suggested to use the intrinsic non-linearity of OM interactions to induce non-reciprocal effects for light~\cite{Manipatruni2009,Hafezi2012}. In the following we describe a related scheme, which allows us to engineer coherent non-reciprocal effects for phonons, where the directionality is simply imposed by the phase difference between two optical driving fields.  

\begin{figure}[t]
\begin{center}
\includegraphics[width=0.85\linewidth]{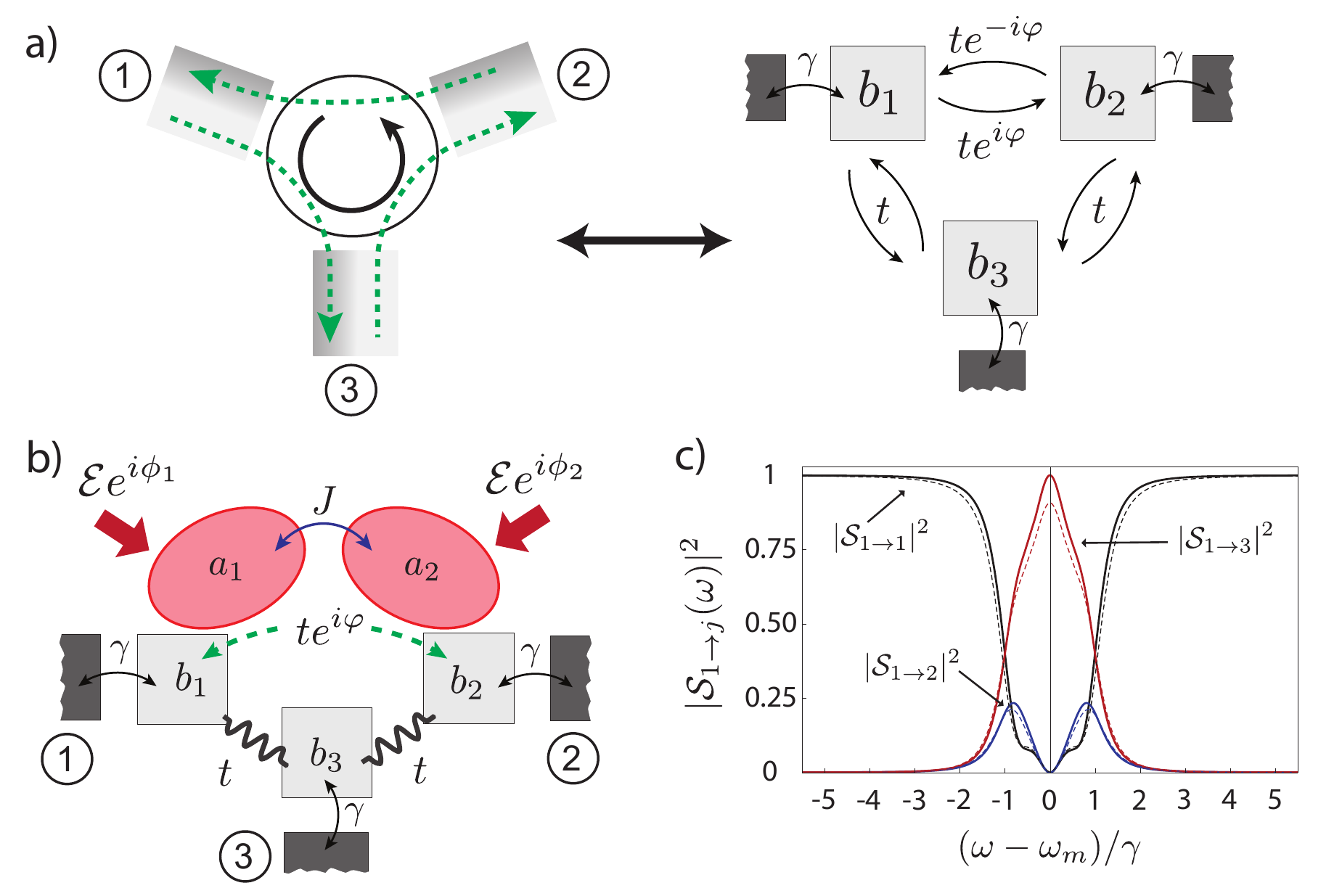}
\caption{A phonon circulator. a) General scheme for realizing a three port circulator based on three coupled phonon cavities, where one of the tunneling amplitudes is complex. A cyclic scattering of phonons between the three ports is achieved for $t= \gamma/2$ and $\varphi=\pi/2$. b) A two mode OM setup for implementing  an effective non-reciprocal tunneling amplitude $te^{i\varphi}$.  The two coupled optical cavities  mediate an effective hopping between the phonon modes $b_1$ and $b_2$, where the overall phase is contorolled by the phases $\phi_{1,2}$ of the external driving fields.   c) Scattering probabilities $|\mathcal{S}_{1\rightarrow j}(\omega)|^2$ into the different ports $j=1,2,3$ of the circulator for an incoming signal of frequency $\omega$ in port 1. The solid line shows the results for an ideal device $(\gamma_0=0)$ and the dashed lines the results for a finite intrinsic loss rate $\gamma_0=\gamma/20$ for each of the three local phonon modes.}
\label{fig:PhononRouter}
\end{center}
\end{figure}

\subsection{A three-port phonon circulator}
To illustrate the basic idea of a non-reciprocal phonon device let us consider the minimal setting of a three port circulator as shown in \fref{fig:PhononRouter}a. The localized phonon modes $b_{j=1,2,3}$ are mutually tunnel-coupled with a strength $|t|$ and coupled to phonon waveguides with a decay rate $\gamma$.  The Hamiltonian is given by
\begin{equation}
H_{\rm circ} =\sum_i \omega_m b^\dag_i b_i + t\left(b_{1}b_{2}^{\dag} e^{i\varphi} +b_{2}b_{3}^{\dag} +b_{3}b_{1}^{\dag} +{\rm H.c.} \right)\;,
\end{equation}
assuming that the resonator frequencies $\omega_m$ are all equal. The crucial term in this setup is a complex tunneling amplitude $t e^{i\varphi}$ between two of the resonators, which cannot be absorbed into local redefinitions of the $b_i$. Therefore, this phase can be associated with an effective magnetic field for the phonon modes. Previously, such a setting has been described for superconducting microwave cavities, where the effective magnetic field can be engineered using superconducting qubits and external fluxes \cite{Koch2010,Nunnenkamp2011}. Below we describe a non-magnetic approach to achieve this complex hopping amplitude in OMS. 

Including the coupling to the waveguides and in a frame rotating with the resonator frequencies $\omega_m$, the QLEs read
\begin{equation}
\left(\begin{array}{c}\dot{b_{1}}\\\dot{b_{2}}\\\dot{b_{3}} \end{array}\right)=-\left(\begin{array}{ccc} \gamma/2 & i t e^{-i\varphi} & i t\\ 
i t e^{i\varphi}&\gamma/2  &i t \\ i t&i t &\gamma/2 \end{array}\right)\left(\begin{array}{c}b_{1}\\b_{2}\\b_{3}\end{array}\right) - \sqrt{\gamma} \left(\begin{array}{c}b_{1,in}\\b_{2,in}\\b_{3,in}\end{array}\right)
\end{equation}
with the input-output relations $b_{j,\rm out}=b_{j,\rm in}+\sqrt{\gamma}b_{j}$. The above set of QLEs can be solved in Fourier space and the input-relation can be applied to obtain the full  scattering matrix $\mathcal{S}(\omega)$ of this system. By choosing $\varphi=\pi/2$ and $t=\gamma/2$, such that the decays into the waveguides are matched to the tunneling amplitudes, we find that for frequencies around $\omega\simeq\omega_{m}$ ($\omega\simeq 0$ in the rotating frame), the scattering matrix is given by  
\begin{equation}\label{eq:ScatteringMatrix}
\left(\begin{array}{c}b_{1,\rm out}\\ b_{2,\rm out}\\ b_{3,\rm out}\end{array}\right)=\left(\begin{array}{ccc}0&1&0\\0&0&i\\i&0&0\end{array}\right)\left(\begin{array}{c}b_{1,\rm in}\\ b_{2,\rm in}\\ b_{3,\rm in}\end{array}\right)\;.
\end{equation}
Up to factors $i$, which can absorbed in the definitions of the operators, Eq. (\ref{eq:ScatteringMatrix}) describes a perfect circulator. For $\varphi=-\pi/2$ we obtain a similar result, but with the scattering direction reversed, i.e. $b_{1,\rm out}=ib_{3,\rm in}$, $b_{2,\rm out}=b_{1,\rm in}$ and $b_{3,\rm out}=ib_{2,\rm in}$. In \fref{fig:PhononRouter}c we plot the scattering probabilities $|\mathcal{S}_{1\rightarrow j}(\omega)|^2$, defined as $b_{j,{\rm out}}(\omega)= \mathcal{S}_{1\rightarrow j}(\omega) b_{\rm 1,in}(\omega)$, as a function of the frequency $\omega$ (relative to $\omega_m$) and for $t=\gamma/2$.  We see the emergence of the ideal circulator relations close to resonance. We also find that these features are robust when we add an additional intrinsic loss rate $\gamma_0=\gamma/20$.

\subsection{Optomechanically engineered non-reciprocity for phonons} 
To implement a complex tunnel amplitude $t e^{i\varphi}$ between two localized mechanical modes we propose a setup as shown in \fref{fig:PhononRouter}b. Here, the localized phonon mode $b_3$ is coupled to modes $b_1$ and $b_2$ mechanically with a (real) tunneling amplitude $t$, while  the tunneling between $b_1$ and $b_2$ is mediated by two driven optical cavities. The Hamiltonian for this system is 
\begin{eqnarray}
H_{\rm circ} &=&\sum_{i=1}^3 \omega_m b_i^\dag b_i 
+t (b_1 b_3^\dag +b_2 b_3^\dag + {\rm H.c.}) + H_c +  H_{\rm om}\;.
\end{eqnarray} 
In the frame rotating with the frequency of an external driving field the Hamiltonian for the two coupled optical cavities is   
\begin{equation}
H_c= \sum_{i=1,2} -\delta_i a_i^\dag a_i  - J (a_1 a_2^\dag + a_1^\dag a_2) +i  \sum_{i=1,2} \mathcal{E}_i \left(a_i^\dag  e^{i\phi_i} - {\rm H.c.}\right) ,   
\end{equation} 
where $\mathcal{E}_i$ are the strengths and $\phi_i$ the phases of the  optical driving fields. 
Finally, 
\begin{equation}
H_{\rm om} =  \sum_{i=1,2} g a_i^\dag a_i (b_i+b_i^\dag), 
\end{equation}
describes the local OM interactions.

As discussed in \sref{sec:NoiseEater}, in the limit of strong driving we can change to a displaced representation and linearize the OM coupling around the classical expectation values $\alpha_i=\langle a_i\rangle$. In the present case the classical field amplitudes are given by
\begin{eqnarray}
\alpha_1&=& \frac{(\kappa -i  \delta_2 )\mathcal{E}_1 e^{i\phi_1}+ i  J \mathcal{E}_2 e^{i\phi_2}}{(\kappa- i\delta_1)(\kappa- i\delta_2) +J^2}, \\
\alpha_2&=&\frac{(\kappa -i  \delta_1 )\mathcal{E}_2 e^{i\phi_2}+ i J \mathcal{E}_1 e^{i\phi_1}}{(\kappa- i\delta_1)(\kappa- i\delta_2) +J^2},
\end{eqnarray}
where we have assumed that both cavities decay with the same rate rate $\kappa$ and we have absorbed non-linear shifts of the detunings by a redefinition of $\delta_i$. We write $\alpha_i=|\alpha_i| e^{i\varphi_i}$  and obtain the linearized OM coupling
\begin{equation}
H_{\rm om} =   \sum_{i=1,2} g |\alpha_i|( e^{-i\varphi_i}a_i   +  e^{i\varphi_i}a_i^\dag) (b_i+b_i^\dag).
\end{equation}
Note that for given $\delta_i$, $J$ and $\kappa$ the external control parameters $\mathcal{E}_i$ and $\phi_i$ provide enough flexibility to independently adjust the phases $\varphi_i$ and keep $|\alpha_1|\approx |\alpha_2|$.   

We focus on the specific case $\delta_i=\delta$, $|\alpha_i|=\alpha$  and introduce symmetric and anti-symmetric optical modes $a_{\pm}=(a_{1}\pm a_{2})/\sqrt{2}$ with detunings $\delta_{\pm}=\delta\pm J$. Further, we assume that $-\delta_{\pm}\approx \omega_{m} \gg g|\alpha_i|$, which allows us to make a rotating wave approximation with respect to $\omega_m$. Then, after changing to a frame rotating with $\omega_m$, the  total Hamiltonian is given by
\begin{eqnarray}
H_{\rm circ}= - \sum_{\nu=\pm} \Delta_\nu   a_\nu^\dag a_\nu   
+t (b_1 b_3^\dag +b_2 b_3^\dag + {\rm H.c.}) +\nonumber\\
\qquad\qquad\frac{g \alpha}{\sqrt{2}}\left[ a^\dag_+(e^{i \varphi_1} b_1+ e^{i\varphi_2} b_2) + a^
\dag_-(e^{i \varphi_1} b_1-e^{i\varphi_2} b_2)  + {\rm H.c.} \right]\;,
\end{eqnarray}
where $\Delta_\nu =\delta_\nu+\omega_m$. In a final step we assume $|\Delta_\nu|\gg g\alpha$ and treat the OM coupling using second-order perturbation theory. Apart from small frequency shifts for the mechanical modes this results in an effective tunneling Hamiltonian
  \begin{equation}
H_{\rm circ}\simeq  t  b_1 b_3^\dag + t b_2 b_3^\dag +  t_{\rm eff} e^{i\varphi} b_{1}b_{2}^{\dag} +    {\rm H.c.} \;, %+ \left( b_{1}b_{2}^{\dag} +b_{1}^\dag b_{2} e^{-i\varphi}\right)\;,
\end{equation}
where $\varphi=\varphi_1-\varphi_2$ and
  \begin{equation}
t_{\rm eff}=\frac{g^2\alpha^2}{2}\left(\frac{1}{\Delta_+}-\frac{1}{\Delta_-}\right)\;.
\end{equation}
Thus, by choosing the external control parameters such that $\varphi=\pm \pi/2$ and  that $t_{\rm eff}$ matches the bare mechanical tunneling amplitude $t$, the setup shown in \fref{fig:PhononRouter}b realizes a switchable three-port phonon circulator as discussed in the previous subsection. Note that the interaction with the optical modes also introduces an additional decay channel with rate $\gamma_{\rm op}\approx g^2\alpha^2\kappa/\bar \Delta^2$, where $\bar \Delta^{-2}=\Delta_+^{-2}+ \Delta_-^{-2}$. 
Compared to the noise filtering scheme described above, this requires slightly lower cavity decays satisfying $\kappa\ll |\bar \Delta| \ll \omega_m$.

As a specific example let us consider the phonon waveguide scenario discussed above with typical phonon frequencies $\omega_m\approx 2\pi\times 4$ GHz and a phonon-waveguide coupling of $\gamma=2\pi\times 25$ MHz.  If we choose $\delta_i=-\omega_m$ we obtain $t_{\rm eff}=g^2\alpha^2/J$ and the conditions $t_{\rm eff}=\gamma/2$ can be achieved for $g\alpha\approx 2\pi\times110$ MHz and $J=2\pi\times 1$ GHz. For the same parameters a value of  $\kappa\leq 2\pi\times 50$ MHz is sufficient to suppress the optically induced decay rate $\gamma_{\rm op}=2 g^2 \alpha^2\kappa/J^2$ below the value $\gamma_{\rm op}/\gamma \leq 0.05$ shown in \fref{fig:PhononRouter}c. This is only slightly below demonstrated values for $\kappa$ in state of the art OM crystal cavities~\cite{Safavi-NaeiniNature2011,Safavi-NaeiniNJP2011}.

\section{Implementations} \label{sec:Implementations}
The fabrication and control of mechanical systems, resonator arrays and
phonon waveguides
as well as their coupling to other microscopic quantum systems (qubits)
is currently a very active field of research. For many of these systems
the general control techniques described in this work could provide the
basis for phonon-based quantum communication applications or mechanical
quantum interfaces in hybrid qubit setups.
In the following we present a brief discussion of several potential
candidate systems for implementing phonon networks.

\subsection{Phonon channels}
As described in  \sref{sec:PhononNetworks},  a 1D phonon channel can be realized by
fabricating a large array of coupled nanomechanical resonators with a
bare frequency $\omega_m$ and nearest neighbor coupling $g< \omega_m$.
This scenario has been experimentally studied, for example in reference~\cite{Buks2002}, where the
resonators were charged up and coupled via electrostatic interactions.
Alternatively, the resonators can also be coupled mechanically via bridges or the support  \cite{Karabalin2009}.  Both
approaches are suitable for realizing phonon channels with frequencies
$\omega_m$ ranging from MHz to a few 100 MHz, where also very high Q-values around $Q\sim 10^5-10^6$ are nowadays routinely achieved.
At  $T\approx 100$ mK this frequency range corresponds to a few tens to a few hundred thermal phonons, which
can still be efficiently suppressed  using the proposed OM noise filtering scheme.

A more practical and very versatile approach for implementing phonon
waveguides based on phononic bandgap materials has recently attracted increasing attention \cite{Olsson2009,Safavi-Naeini2010}. Here, a 2D structure with periodically varying mechanical properties is designed such that a complete band gap in the mechanical dispersion relation appears.
Then, phonon wave guides can be realized by introducing line-defects in these structures, to which the phonons are confined as long as their frequency lies within the bandgap of the bulk phononic crystal. With this
approach phonon waveguides with frequencies $\omega_m \approx 1-10$ GHz can be realized, where even at $T=4$ K the thermal occupation of the waveguide is only a few tens of quanta.  Further, as indicated in Figure \ref{fig:PhononNetwork}b, such waveguides can be easily combined with
localized phonon cavities and integrated OM devices \cite{Safavi-NaeiniNJP2011}.

\subsection{Qubit-phonon interfaces}
The implementation of phonon quantum
networks also  requires the realization of coherent and controllable
interfaces between mechanics and stationary qubits.  Here the ability of
mechanical systems to respond to  weak optical, electrical and magnetic
forces enables the coupling of mechanical resonators to a large variety
of spin or charge based qubits and a few prototype examples are outlined
in the following.

\subsubsection{Spin qubits.}  Qubits encoded in localized spin degrees
of freedom, for example in spins associated with  Nitrogen vacancy (NV)
impurities in diamond \cite{Jelezko2006} or phosphor donors in silicon \cite{Tyryshkin2012}, can be coupled to mechanical motion using magnetized tips \cite{Arcizet2011,Kolkowitz2012, Mamin2007}. A strong magnetic field
gradient $\nabla B$ from the tip induces a position dependent Zeeman shift of the spin resulting in an interaction of
the form
\begin{equation}\label{eq:BareSpinCoupling}
H_{\rm node}= \frac{\omega_0}{2} \sigma_z + \omega_m b^\dag b +  \lambda
(b+b^\dag)\sigma_z + \frac{\Omega}{2}\cos(\omega_{mw} t)  \sigma_x.
\end{equation}
Here $\omega_0\sim $ GHz is the bare spin splitting and $\lambda= g_s
\mu_B a_0 \nabla B/(2\hbar) $ is the Zeeman shift per zero point motion
$a_0$, where $g_s\simeq 2$ and $\mu_B$ is the Bohr magneton. For
nano-scale mechanical resonators with frequencies in the few MHz regime
this coupling can be substantial and reach values  $\lambda/(2\pi)\approx
10-100$ kHz \cite{Rabl2009}.  Base on this coupling the use of mechanical transducers to mediate spin-spin
interactions in small resonator arrays has been proposed previously \cite{Rabl2010},
and the present techniques can be used to extend these ideas to larger
networks. To achieve an effective Jaynes-Cummings type interaction as given in Eq.
(\ref{eq:Hnode}), the spin is driven by a near resonant microwave field with frequency
$\omega_{\rm mw} =\omega_0+\delta$ and Rabi frequency $\Omega$ as described
by the last term in Eq. (\ref{eq:BareSpinCoupling}). Then, by changing
into a dressed spin basis  and making a rotating wave approximation with respect to $\omega_0$, the
effective interaction is given by \cite{Stannigel2011,Rabl2009}
\begin{equation}
H_{\rm node}\simeq \frac{\Delta_q}{2} \tilde \sigma_z + \omega_m b^\dag
b +  \lambda \sin(\theta) (\tilde \sigma_+ b + \tilde \sigma_- b^\dag),
\end{equation}
where $\Delta_q=\sqrt{\delta^2+\Omega^2}\sim \omega_m$ and
$\tan(\theta)=\Omega/\delta$. The qubit-resonator coupling can be
controlled by adiabatically varying the effective detuning $\Delta_q$ or
the mixing angle $\theta$.

Instead of using external magnetic field gradients, alternative schemes to strongly couple spins to mechanical motion 
have been recently suggested for carbon nanotubes \cite{Palyi2011,Ohm2011}. Here a single electron is localized on the nanotube and couples to its vibration 
via spin-orbit interactions.  In this case even larger couplings $\lambda/(2\pi) \approx 0.5$ MHz and
larger mechanical frequencies $>100$ MHz are achievable. Although a controlled fabrication and positioning of carbon nanotubes is still challenging, a phonon quantum bus could be realized by electrically coupling the nanotube to other mechanical resonators, which can be fabricated in a more controlled manner.

\subsubsection{Superconducting qubits.} The coupling of nanomechanical
systems to various types of superconducting qubits has been demonstrated in recent experiments \cite{OConnell2010,LaHaye2009,Etaki2008}. While for superconducting qubits other ways for communication exist,
for example via microwave transmission lines, the coupling to
phononic channels could still be interesting for creating interfaces to other quantum systems, especially to optical qubits for long-distance quantum communication \cite{Stannigel2010,Safavi-NaeiniNJP2011}. Depending on the type of qubit (`charge', `phase', `flux', ...) the qubit-resonator interaction can
be due to electrostatic \cite{Armour2002,Martin2004}, piezo-electric \cite{OConnell2010} or magnetic interactions \cite{Zhou2006,Xue2007a,Jaehne2008},
and can in all three cases be substantially larger than in the case of spin qubits. Since the bare transition frequency of superconducting qubits is typically in the GHz regime, a resonant coupling to mechanical systems in the MHz range can again be achieved as described above, namely by applying additional driving fields to compensate for the frequency mismatch \cite{Martin2004,Rabl2004}. Alternatively, superconducting qubits could be coupled directly to high frequency mechanical modes as demonstrated in Ref.~\cite{OConnell2010}. 

\subsubsection{Quantum dots and defect centers.}  The bending of a
mechanical resonator locally deforms  the lattice configuration of the
host material and by that shifts the electronic states associated with
artificial quantum dots or naturally occurring defect centers in solids.
In the bulk this deformation potential interaction is usually responsible for line
broadening of optical transitions and other decoherence effects,  but in
the case of confined  modes it can also lead to a significant
coherent coupling to single phonons. In Ref. \cite{WilsonRae2004} the coupling of a
quantum dot to the fundamental bending mode of a doubly clamped beam has
been considered, leading to a deformation potential coupling of the form
\begin{equation}\label{eq:Hdef}
H_{\rm def} = \lambda (b+b^\dag)|e\rangle\langle e|, \qquad \hbar
\lambda \approx (\Xi_e-\Xi_g) \frac{a_0  t}{2l^2},
\end{equation}
where $|e\rangle$ denotes the electronically excited state, $l$ the beam
length, $t$ its thickness and $a_0$ the zero point motion. The $\Xi_e$
and $\Xi_g$ are deformation potential constants for the ground and
excited electronic states and for typical values $\Xi_{e,g}\sim 1-10$
eV  and micron sized beams a coupling strength of $\lambda/2\pi\approx
10$ MHz can be achieved. This is already comparable to the radiative
lifetime $\Gamma_e$ of the electronically excited state, but can in principle be pushed  to values  $\lambda\sim 1$ GHz
using much smaller, phononic Bragg-cavities as suggested in Ref.~\cite{Soykal2011}.

As a potential scenario where the deformation potential interactions could 
be used to implement a controllable qubit phonon interface, we consider
an NV center embedded in a diamond nanoresonator. The electronic ground
state of the NV defect is a spin triplet, where two states $|0\rangle =|m_s= -1\rangle$ and $|1\rangle=|m_s=1\rangle$  can be used
to encode quantum information.
The qubit states in the electronic ground state, which are highly immune against external perturbations, can be coupled via an optical  Raman process involving
an electronically excited state, which, in contrast, is strongly coupled to phonons \cite{Maze2011}.
In \ref{app:NVPhonon} we adiabatically eliminate the dynamics of the excited state and derive an effective spin-phonon interaction of the form
\begin{equation}\label{eq:EffectiveSpinPhonon}
H_{\rm eff}\simeq   \lambda_{\rm eff}(t)  (\sigma_+ b + \sigma_- b^\dag).
\end{equation}
Here the $\sigma_\pm$ are Pauli operators for the qubit subspace and for optimized laser detunings  $ \lambda_{\rm eff}(t)=4 \lambda \Omega_0(t)\Omega_1(t)/\omega_m^2$ is a tunable coupling where $\Omega_{0,1}(t)$ are the optical Rabi frequencies.  Note
that this coherent interaction is accompanied by dissipation at an average rate $\bar 
\Gamma_{\rm eff}(t) = 4\lambda \Omega_0(t)\Omega_1(t)/\omega_m^2$ and the ratio
$\lambda_{\rm eff} /\bar \Gamma_{\rm eff} =\lambda/\Gamma$ is not affected by the
off-resonant Raman coupling. This is in contrast to cavity QED where the optical field couples to
the atomic coherence and not to the population of the excited state as described by $H_{\rm def}$.

\section{Conclusions and outlook} \label{sec:Conclusions}

In summary we have described the implementation of dissipative as well
as coherent OM control elements for realizing quantum communication
protocols in extended phononic networks. We have shown how OM continuous
mode cooling schemes can be used to create a cold frequency window in
the thermal noise spectrum of a phononic channel and we have analyzed
the fidelity of quantum state transfer protocols under these conditions.
Further, we have proposed the realization of non-reciprocal phononic
elements, which rely on strong coherent OM interactions and where the
directionality is simply  controlled by the phase of an external laser
field. Based on this principle, various switches and routers for
propagating phonons can be constructed, which allow for the
implementation of efficient quantum communication protocols also in
larger phonon networks.

Both the OM noise filter as well as the phonon router can be realized
with state-of-the-art mechanical systems. Combined with the ongoing
developments in the control of qubit-resonator interactions, they could soon
provide the elemental tools for implementing phononic quantum
networks.  As potential applications of such networks we envision the
distribution of entanglement within a larger quantum computing
architecture, where propagating phonons can replace or complement direct
qubit shuttling techniques \cite{Kielpinski2002,Taylor2005}. Here  the ability of mechanical systems
to interact with various different types of qubits, makes phononic
channels particularly suited for hybrid qubit settings. Beyond
quantum communication applications, the OM control techniques described here may also be used to probe the propagation and scattering of single-phonon wave packets.

\section*{Acknowledgments}

The authors would like to thank S. Bennett, F. Marquardt, O. Painter, A. Safavi-Naeini  for stimulating  discussions. This work
was supported by the EU network AQUTE and the Austrian Science Fund (FWF)
through SFB FOQUS and the START grant Y 591-N16. M.L. acknowledges support by NSF, CUA, DARPA and the Packard Foundation.

\appendix

\section{Coupled resonator arrays}\label{app:ResonatorArray}
In this appendix we derive the effective propagation equations and input-output relations for phononic quantum channels consisting of a large array of $N_{\rm ch}\gg1 $ coupled mechanical resonators. Assuming a homogenous system the resonator array is described by the Hamiltonian 
\begin{equation}
H_{\rm channel}= \sum_{\ell=1}^{N_{\rm ch}} \left( \frac{p_\ell^2}{2m} +\frac{1}{2} m \omega_0^2 x_\ell^2 \right)+ \frac{k}{2}\sum_{\ell=1}^{N_{\rm ch}-1} (x_\ell-x_{\ell+1})^2,
\end{equation}
and the equations of motion for the position operators $x_\ell$ are  given by
\begin{equation}\label{eq:ddotxn}
\ddot x_\ell = -  \omega_0^2 x_\ell - \omega_0K ( 2x_\ell - x_{\ell+1} - x_{\ell-1}), 
\end{equation}
where $K= k/(m\omega_0)$ is the nearest neighbor phonon-phonon coupling strength. For a large array we can assume periodic boundary conditions and write  
\begin{equation}
x_\ell(t) = \frac{1}{\sqrt{N_{\rm ch}}} \sum_{n}  \sqrt{\frac{\hbar}{2m\omega_n}} \left( e^{i 2\pi n \ell /N_{\rm ch}}  b_n(t) 
 + e^{-i 2\pi n \ell /N_{\rm ch}} b_n^\dag (t)\right),   
  \end{equation}
where $b_n(t)=b_n e^{-i\omega_n t}$ are bosonic operators for the plane wave modes labeled by $n=-(N_{\rm ch}/2-1), \dots, N_{\rm ch}/2$ and normalized to $[b_n,b_{n'}^\dag]=\delta_{n,n'}$. The eigenfrequencies are given by 
\begin{equation}
\omega_n=  \sqrt{ \omega_0^2  +2  K  \omega_0 \left[1-  \cos(2\pi n /N_{\rm ch}) \right] }\,.
\end{equation}
We are interested in the regime where the total length $L$ as well as the other relevant scales of the network are large compared to the spacing $a$ between the individual resonators.  
We introduce a continuous field 
\begin{equation}
\Xi(z)= \frac{1}{\sqrt{N_{\rm ch}}} \sum_{q}  \sqrt{\frac{\hbar}{2m\omega_q}} \left( e^{i q z}  b_q 
 + e^{-i k z} b_k^\dag \right), 
\end{equation}
such that $x_\ell= \Xi(z=a \ell) $. The quasi-momentum $q$ is restricted to the first Brillouin zone  $q\in (-\pi/a,\pi/a]$ and $\omega_q=  \sqrt{ \omega_0^2  +2  K  \omega_0 \left[1-  \cos(qa) \right] }$.

To proceed we now consider the tight binding limit $K\ll \omega_0$, where $\omega_q \simeq  \omega_0 + K(1-\cos(qa))$ and the phonon modes form a band of width $\Delta \omega= 2K$ around a large center frequency  $\bar \omega=\omega_0+K$.  Further, for frequencies around $\bar \omega$  the dispersion relation is approximately linear and can be written as   
\begin{equation}
\omega_q \simeq  \bar\omega + c(|q|- \pi/(2a)) =\tilde \omega_0 +c |q|,
\end{equation}
where $c=K a$  is the sound velocity and $\tilde \omega_0=\omega_0-(\pi/2-1)K$ a frequency offset. Therefore, as long as we are interested in the dynamics of phonon modes    away from the band edges we can set $H_{\rm channel}\simeq \sum_q (\tilde \omega_0 +c |q|) b_q^\dag b_q$  and approximate the displacement field by 
\begin{equation}
\Xi(z)\simeq \frac{\bar x_0}{\sqrt{\Delta \omega}} \left[ \Phi(z) + \Phi^\dag(z)  \right].
\end{equation}
Here  $\bar x_0= \sqrt{\hbar/2m\bar \omega}$ and the field $\Phi (z) =\sqrt{2c/L} \sum_{q}  e^{ iq z} b_q$ is normalized to 
\begin{eqnarray}
\left[ \Phi(z,t), \Phi^\dag(z',t')\right] &\simeq&  e^{-i\tilde \omega_0 (t-t')}\left[ \delta \left( \Delta t - \frac{\Delta z}{c}\right)+\delta \left( \Delta t+ \frac{\Delta z}{c}\right)\right],
\end{eqnarray} 
where $\Delta t=t-t'$ and $\Delta z=z-z'$.  The field operators can be decomposed into a right- and a left-moving component  $\Phi(z)= \Phi_R(z)+\Phi_L(z)$, as defined in equations \eref{eq:PhiLR} and \eref{eq:PhiLRCommutator}.

The coupling of the localized phonon modes to the phonon channel can be written as 
\begin{equation}
H_{\rm int}= \frac{1}{2} \sum_{i=1}^N \sum_\ell k_{i,\ell}  (x_i-x_{\ell})^2,
\end{equation}
where $k_{i,\ell}=k_{\rm loc}$ is non-zero only for site $\ell$, which are next to a local mode $i$ (we can assume side-coupled resonators, such there is only one neighbor). We define $K_{\rm loc}=k_{\rm loc} x_0 \bar x_0$ where $x_0$ the zero point motion of the local mode. Then, under the assumption that $\omega_m\approx \bar \omega \gg K_{\rm loc}$, we can make a RWA and obtain the coupling given in equation \eref{eq:Hint} with a  resonator decay rate given by   $\gamma=2K_{\rm loc}^2/\Delta \omega$.
 
\subsection{Propagation losses} \label{app:ResonatorArray:Losses} 
To model propagation losses in the phonon waveguide we add an intrinsic loss channel with rate $\gamma_0$ for each of the waveguide modes. Making the RWA right from the beginning, the dynamics of the whole resonator array is modeled by the coupled QLEs, 
\begin{equation}
\dot b_\ell= -\left(i(\omega_0+K)+\frac{\gamma_0}{2}\right)  b_\ell +  \frac{iK}{2} \left(b_{\ell-1} + b_{\ell +1}\right) -\sqrt{\gamma_0} b_{\ell,{\rm in}}(t),
\end{equation} 
where $b_\ell$ is the bosonic operator for the mechanical resonator at site $\ell$ and the $b_{\ell,{\rm in}}(t)$ are uncorrelated thermal noise operators $\langle b^\dag_{\ell,{\rm in}}(t)b_{\ell',{\rm in}}(t')\rangle =N_{\rm th}\delta_{\ell,\ell'} \delta(t-t')$. In the corresponding momentum representation $b_q=1/\sqrt{N_{\rm ch}} \sum_\ell e^{iqa\ell} b_\ell$ and $b_{q,{\rm in}}=1/\sqrt{N_{\rm ch}} \sum_\ell e^{iqa\ell} b_{\ell,{\rm in}}$, the coupling is diagonal
 \begin{equation}
\dot b_q= -\left(i\omega_q+\frac{\gamma_0}{2}\right)  b_q -\sqrt{\gamma_0} b_{q,{\rm in}}(t),
\end{equation} 
where as above $\omega_q=\omega_0+K(1-\cos(qa))\approx \tilde \omega_0 + c |q|$. From this equation we obtain the propagation equation for the right moving field $\Phi_R(z,t)=\sqrt{2c/L}\sum_{q>0} e^{iqz} b_q(t)$, 
\begin{equation}
\left(\frac{\partial}{\partial t}+ c \frac{\partial}{\partial z}\right) \Phi_R(z,t) = -\left(i\tilde \omega_0 + \frac{\gamma_{0}}{2}\right)   \Phi_R(z,t) - \sqrt{\gamma_{0} c}\,  \Phi_{\rm th}(z,t).
\end{equation}
Here the thermal noise field is defined as 
\begin{equation}
 \Phi_{\rm th}(z,t)= \sqrt{\frac{2}{L}}\sum_{q>0} e^{iqz} b_{q,{\rm in}}(t),
\end{equation}
such that it  is normalized $\langle   \Phi^\dag_{\rm th}(z,t) \Phi_{\rm th}(z',t')\rangle =N_{\rm th} \delta(z-z')\delta(t-t')$.

\section{Optomechanical cooling}\label{app:OMCooling} 
After linearizing the OM coupling the QLEs (\ref{eq:QLE_Single_a}) and (\ref{eq:QLE_Single_b}) can be written as 
\begin{equation}
\partial_t \vec A(t)= -\mathcal{M} \vec A(t)- \sqrt{2\kappa}   \vec A_{\rm in}(t)  -\sqrt{\gamma_0}   \vec B_{0,{\rm in}}(t),   
\end{equation}   
where we have grouped operators as $\vec A(t)=(a(t),a^\dag(t),b(t),b^\dag(t))^T$,
 $\vec A_{\rm in}(t)= (a_{\rm in} (t) ,a_{\rm in}^\dag (t), 0,0)^T$ and $\vec B_{0,{\rm in}}(t)  = (0,0,  b_{0,{\rm in}} (t) ,b_{0,{\rm in}}^\dag (t))^T$.
The matrix $\mathcal{M}$ is given by
\begin{equation}
\mathcal{M}=
\left(\begin{array}{cccc}
-i\delta + \kappa     &   0     & i g\alpha         &   i g\alpha  \\
0           & i\delta + \kappa   &- i g\alpha^*& - i g\alpha^*      \\
i g \alpha^*           & i g \alpha & i\omega_m +\frac{\gamma_0}{2}& 0      \\
-i g \alpha& -i g \alpha^*    & 0    & -i\omega_m +\frac{\gamma_0}{2}
\end{array}\right).
\end{equation}
We define the Fourier-transformed operators as $a(\omega)=\frac{1}{\sqrt{2\pi}}\int dt\ e^{i\omega t}a(t)$, $a^\dag(\omega)=[a(\omega)]^\dag$ and obtain
\begin{equation}\label{eq:AppAw}
 \vec A(\omega)= -\mathcal{X}(\omega)\left( \sqrt{2\kappa}   \vec A_{\rm in}(\omega)  +\sqrt{\gamma_0}   \vec B_{0,{\rm in}}(\omega)\right),  
\end{equation} 
where we set $ \mathcal{X}(\omega) =\left[ \mathcal{M}-i\omega \mathbbm{1} \right]^{-1}$
and  now  $\vec A(\omega)=(a(\omega),a^\dag(-\omega),b(\omega),b^\dag(-\omega))^T$, etc. The non-vanishing correlations of the noise operators are $\langle a_{\rm in} (\omega) a^\dag_{\rm in}(\omega')\rangle=\delta (\omega-\omega')$,  $\langle b_{\rm 0, in}(\omega) b^\dag_{\rm 0, in}(\omega')\rangle=(N_{\rm th}+1) \delta (\omega-\omega')$ and $\langle b^\dag_{\rm 0, in}(\omega) b_{\rm 0, in}(\omega')\rangle=N_{\rm th} \delta (\omega-\omega')$,  within the relevant frequency range.  Then,  the stationary fluctuation spectrum of the mechanical mode is given by
\begin{equation}
\langle  b^\dag (\omega) b(\omega')\rangle=\left[ 2\kappa\left| \mathcal{X}(\omega)_{32}\right|^2 \!+ \! \gamma_0 N_{\rm th} \left( \left| \mathcal{X}(\omega)_{33}\right|^2 \! +\!\left|\mathcal{X}(\omega)_{34}\right|^2 \right) \right] \delta(\omega-\omega').
\end{equation}
Under the weak coupling and sideband resolved condition this expression simplifies to the result given in equation \eref{eq:SingleModeCoolingResult}.

In \sref{sec:NoiseFilter} we are interested in spectrum of the scattered field $b_{\rm out}(t)$ in the case where the OMS is in addition coupled to the phonon waveguide with rate $\gamma$.   We define $\vec A(\omega)$ and the intrinsic noise $\vec B_{\rm 0,in}(\omega)$ as above, but we set $\vec A_{\nu={\rm in,out}}(\omega)=(a_{\nu} (\omega),a_{\nu}^\dag(-\omega),b_{\nu}(\omega),b_{\nu}^\dag(-\omega))^T$. The input-output relations can then be written as $\vec A_{\rm out}(\omega)=\vec A_{\rm in}(\omega)+ \sqrt{\mathcal{R}} \vec A(\omega) $ where $\mathcal{R}={\rm diag}(2\kappa,2\kappa,\gamma, \gamma)$ is a diagonal matrix. Then, together with equation \eref{eq:AppAw} we obtain
\begin{equation}
\vec A_{\rm out}(\omega)= \mathcal{S}(\omega) \vec A_{\rm in}(\omega)-  \mathcal{S}^\prime(\omega)  \vec B_{0,{\rm in}}(\omega).
\end{equation}
where $\mathcal{S}(\omega)=\mathbbm{1}-\mathcal{R} \mathcal{X}(\omega)$ and $\mathcal{S}^\prime(\omega)=\sqrt{\gamma_0\mathcal{R}}\mathcal{X}(\omega)$
and in the definition of $\mathcal{X}(\omega)$ we replaced $\gamma_0\rightarrow \gamma_0+\gamma$. The filtered noise spectrum of the out-field is given by 
\begin{equation}
N_F(\omega)=|\mathcal{S}(\omega)_{32}|^2+ N_{\rm th}( |\mathcal{S}(\omega)_{33}|^2 +|\mathcal{S}(\omega)_{34}|^2+|\mathcal{S}^\prime(\omega)_{33}|^2 +|\mathcal{S}^\prime(\omega)_{34}|^2),
\end{equation}
where in equation \eref{eq:NFideal} the approximation $N_F(\omega)\simeq N_{\rm th} |\mathcal{S}(\omega)_{33}|^2$ has been made.

\subsection{Multi-mode cooling} 
We can use the same approach to solve for the stationary state of the multimode system described in \sref{sec:MultiMode}. In the following we assume for simplicity the validity of the RWA, such that the QLEs for the equations for the annihilation operators $\vec A=(a,b_1,\dots b_N)^T$ form a closed set
\begin{equation}
\partial_t \vec A(t)= -\mathcal{M} \vec A(t)- \sqrt{2\kappa}   \vec A_{\rm in}(t)  -\sqrt{\gamma_0}   \vec B_{0,{\rm in}}(t), 
\end{equation}   
where now $\vec A_{\rm in}(t)= (a_{\rm in} (t) ,0,\dots, 0)^T$ and  $\vec B_{0,{\rm in}}(t)  = (0, b^{(1)}_{0,{\rm in}} (t) ,\dots, b^{(N)}_{0,{\rm in}} (t))^T$ and the matrix $\mathcal{M}$ can be derived from equations \eref{eq:QLE_Multi_a}-\eref{eq:QLE_Multi_bN}. 
Following the same steps as above we obtain 
\begin{equation}
\langle  b_i^\dag (\omega) b_i(\omega')\rangle=  \gamma_0 N_{\rm th}  \left( \sum_{j=1}^{N} \left|  \mathcal{X}(\omega)_{i+1,j+1 }\right|^2\right)\delta(\omega-\omega'),
\end{equation}
which we have used to evaluate the multi-mode cooling results presented in \fref{fig:Multimode}c.

\section{Cascaded master equation}\label{app:CascadedME}
To simulate the state transfer between two nodes in the presence of incident noise and OM cooling, we map the qubit QLEs (\ref{eq:qubitQLE1}) and (\ref{eq:qubitQLE2}) with time dependent decay rates $\Gamma_{1}(t)$ and $\Gamma_{2}(t)$ onto an equivalent cascaded master equation \cite{QuantumNoise}, which we can then integrate numerically. Since we cannot treat the filtered input noise directly using a master equation we consider a unidirectional network where the OM cooled phonon cavity is included as a first system. The OM cooling is modeled by an additional decay channel for the phonon cavity at a rate $\gamma_{\rm op}=2g^{2}|\alpha|^2/\kappa$, which is matched to the decay rate $\gamma$ of the phonon cavity into the waveguide. The cascaded master equation, which describes this system is given by
\begin{equation}
\dot{\rho}=- i[H,\rho]+(N_{\rm th}+1)\mathcal{D}[S]\rho+N_{\rm th}\mathcal{D}[S^{\dag}]\rho+\gamma_{\rm op}\mathcal{D}[b]\rho\;,
\end{equation}
where $\mathcal{D}[c]\rho= c\rho c^\dag -(c^\dag c \rho +\rho c^\dag c)/2$. By identifying $c_0\equiv b$, $c_{j=1,2}=\sigma_-^{j}$ and $\Gamma_{0}=\gamma$ the collective operator $S=\sum_{k=0,1,2}\sqrt{\Gamma_{k}}c_{k}$ and  $H=-\frac{i}{2} \sum_{k>l}\sqrt{\Gamma_{k}\Gamma_{l}}(c^{\dag}_{k}c_{l}-c^{\dag}_{l}c_{k})$.  $N_{\rm th}$ is the thermal occupation number of the incident white noise.

\section{NV-phonon coupling}\label{app:NVPhonon}
In this appendix we show how for an NV center in diamond the deformation potential coupling can be used to realize the effective spin-phonon interaction (\ref{eq:EffectiveSpinPhonon}). The NV center has $S=1$ triplet ground state and we assume that the qubit is encoded in the states $|0\rangle\equiv |m_s=-1\rangle$ and $|1\rangle=|m_s=+1\rangle$. The qubit states are coupled by external laser fields to an electronically excited state $|e\rangle$, which is coupled to the deformation of the local lattice structure induced by the vibration of the beam. In the frame rotating with the laser frequencies the Hamiltonian is given by
\begin{equation}
\label{HNV}
H=\sum_{j=0,1}\left[\Delta_{j}|j\rangle\langle j|+\Omega_{j}(t)(|j\rangle \langle e|+|e\rangle\langle j|)\right]+\omega_{m}b^{\dag}b+\lambda|e\rangle\langle e|(b+b^{\dag})\;,
\end{equation}
where the $\Omega_j(t)$ are tunable Rabi frequencies and we have chosen the zero of energy to coincide with the excited state level such that $\Delta_{0}$ and $\Delta_{1 }$ are detunings of the drive lasers from the $|0\rangle$ to $|e\rangle$ and $|1\rangle$ to $|e\rangle$ transitions, respectively. The interaction term $\sim \lambda$ can be eliminated by a \textit{polaron transformation}
\begin{equation}
H\rightarrow e^{S}He^{-S}\qquad\mathrm{with}\qquad S=\frac{\lambda}{\omega_{m}}|e\rangle\langle e|(b^\dag -b)\;,
\end{equation}
which transforms the Hamiltonian into
\begin{equation}
H= \omega_{m}b^{\dag}b+ \sum_{j=0,1}\Delta_{j}|j\rangle\langle j|+ \left[\left(\Omega_{0}|0\rangle \langle e|+ \Omega_{1}|1\rangle\langle e|\right)A +{\rm H.c.} \right]
\end{equation}
and is still exact. Since the ratio $\eta\equiv\lambda/\omega_{m}\ll1$, we can expand to first order in $\eta$ and obtain $A\simeq 1-\eta (b^\dag-b)$.

Our goal is to induce a coherent Raman transition from state $|0\rangle$ to state $|1\rangle$ while simultaneously absorbing a phonon from the nanomechanical oscillator.
To make this process resonant we set $\Delta_{0}=\Delta+\omega_m/2$ and $\Delta_1=\Delta-\omega_m/2$, where $\Delta=(\Delta_{0}+\Delta_{1})/2$ is the overall detuning. We change into an interaction picture with respect to $H_0=  \omega_{m}b^{\dag}b+ \sum_{j=0,1}\Delta_{j}|j\rangle\langle j|$ and write the total state as $|\tilde \psi\rangle(t)= \sum_{j=0,1,e} |\tilde \psi_j\rangle(t)$. 
The equations of motion are then given by
\begin{eqnarray} 
\fl \qquad \partial_t |\tilde \psi_e\rangle (t) &=& - \frac{\Gamma_e}{2}|\tilde\psi_e\rangle(t)-i \sum_{j=0,1} \Omega_j e^{-i\Delta_j t}\left(1+   \eta  e^{i\omega_m t} b^\dag - \eta  e^{-i\omega_m t} b \right) |\tilde \psi_j\rangle (t)\\
\fl \qquad\partial_t |\tilde \psi_{j=0,1}\rangle (t) &=& -i \Omega_j e^{i\Delta_j t}\left(1-  \eta  e^{i\omega_mt} b^\dag + \eta  e^{-i\omega_mt} b \right) |\tilde \psi_e\rangle (t), 
\end{eqnarray}
where we have added an imaginary part to model the radiative decay of the exited state with rate $\Gamma_e$ \footnote{For simplicity we here omit the associated recycling terms, which can be included, for example, using a stochastic wavefunction formalism~\cite{QuantumNoise}.}. Assuming that $|\Delta_{0,1}| \gg \Omega_{0,1}$, we can approximately integrate the equation of motion for $|\tilde \psi_e\rangle(t)$,   
\begin{equation}
\fl \qquad |\tilde \psi_e\rangle(t) \simeq  \sum_{j=0,1} \Omega_j \left( \frac{e^{-i\Delta_j t} }{\Delta_j+i\frac{\Gamma_e}{2}} - \frac{\eta b e^{-i(\Delta_j+\omega_m) t}  }{(\Delta_j+\omega_m)+i\frac{\Gamma_e}{2} }+ \frac{\eta b^\dag e^{-i(\Delta_j-\omega_m) t} }{(\Delta_j+\omega_m)+i\frac{\Gamma_e}{2} }\right) |\tilde \psi_j\rangle(t),
\end{equation}
and insert this result back into the equations of  motion for $|\tilde \psi_{0,1}\rangle(t)$. By keeping only non-rotating terms, we obtain for the $|\tilde \psi_0\rangle$ subspace,  
\begin{equation}
\fl \qquad \partial_t |\tilde \psi_0\rangle (t)= -  \frac{i\Omega_0^2}{\Delta_0+i\frac{\Gamma_e}{2}} |\tilde \psi_0\rangle(t)  -i  \left[ \frac{ \eta \Omega_0\Omega_1}{\Delta_1+i\frac{\Gamma_e}{2} } - \frac{\eta \Omega_0\Omega_1}{(\Delta_1+\omega_m)+i\frac{\Gamma_e}{2} } \right]  b |\tilde \psi_1\rangle(t), 
\end{equation}
and a similar result for the evolution of $|\tilde \psi_1\rangle(t)$. In the limit $|\Delta_j|,|\Delta_j\pm \omega_m| \gg \Gamma_e$ and using the definition $\Delta_{0,1}=\Delta\pm \omega_m/2$, we  can identify an effective spin-phonon coupling $\lambda_{\rm eff}$ and effective decay rates $\Gamma^{(j)}_{\rm eff}$ for each spin level,
\begin{equation}
\lambda_{\rm eff} = \frac{\lambda \Omega_0\Omega_1}{\Delta^2-\omega_m^2/4},\qquad \Gamma^{(j)}_{\rm eff}=  \Gamma_e \frac{\Omega_j^2}{\Delta_j^2}.
\end{equation} 
The ratio between the coherent coupling and the mean decay rate $\bar \Gamma_{\rm eff}=(\Gamma^{(0)}_{\rm eff}+\Gamma^{(1)}_{\rm eff})/2$ is optimized for $\Delta=0$ where $\lambda_{\rm eff}/\bar \Gamma_{\rm eff}=\lambda/\Gamma_e$.

Here we have assumed that the NV center is coupled to a single vibrational mode. In general, nanostructured resonators will support multiple mechanical modes so that the phononic part in the Hamiltonian (\ref{HNV}) generalizes to $H_{\rm phonon}=\sum_{k}\omega_{k}b_{k}^{\dag}b_{k}+|e\rangle\langle e|\sum_{k}\lambda_{k}(b_{k}+b_{k}^{\dag})$. However, since the higher-frequency modes of nanostructured resonators are separated from the fundamental mode by $\sim$ GHz, the other modes are highly off-resonant and their contributions to the resulting spin-phonon coupling are negligibly small. See reference \cite{WilsonRae2004} for a similar discussion.

\section*{References}
\bibliographystyle{unsrt}
\bibliography{references}

\end{document}